\RequirePackage{amsthm}
\documentclass[pdflatex,sn-mathphys-num, referee]{sn-jnl}

\usepackage{graphicx}%
\usepackage{multirow}%
\usepackage{amsmath,amssymb,amsfonts}%
\usepackage{upgreek}
\usepackage{mathrsfs}%
\usepackage{mathtools}
\usepackage[title]{appendix}%
\usepackage{xcolor}%
\usepackage{textcomp}%
\usepackage{manyfoot}%
\usepackage{booktabs}%
\usepackage{algorithm}%
\usepackage{algorithmicx}%
\usepackage{algpseudocode}%
\usepackage{listings}%
\usepackage{subfiles}
\usepackage{setspace}
\usepackage{makecell}
\usepackage{lineno}
\usepackage{longtable}   
\usepackage{adjustbox}   
\usepackage{array}
\usepackage{float}
\usepackage{placeins}
\usepackage{xfrac}
\usepackage{changepage}
\usepackage{jabbrv}

\theoremstyle{thmstyleone}%
\theoremstyle{thmstyletwo}%

\theoremstyle{thmstylethree}%

\usepackage{calc}

\makeatletter
\large
\newlength\my@tdima
\newlength\my@tdimb
\newcounter{my@tcnta}
\newcounter{my@tcntb}

\@settopoint\topmargin

\@settopoint\textwidth
\@settopoint\oddsidemargin
\@settopoint\evensidemargin

\@settopoint\marginparwidth
\makeatother

\begin{document}

\title[Article Title]{One-Dimensional Modeling of Blood Flow: A Comprehensive Yet Concise Review}


\author*[1]{\fnm{Daehyun} \sur{Kim}}\email{kimx4718@umn.edu}

\author*[1]{\fnm{Jeffrey} \sur{Tithof}}\email{tithof@umn.edu}

\affil[1]{\orgdiv{Department of Mechanical Engineering}, \orgname{University of Minnesota}, \orgaddress{\street{111 Church St SE}, \city{Minneapolis}, \state{MN}, \postcode{55455}, \country{United States}}}


\abstract{
One-dimensional (1D) blood flow simulations are extensively used in cardiovascular research due to their computational efficiency and effectiveness in analyzing pulse wave dynamics. Despite their versatility and simplicity, there is a lack of a unified, step-by-step guide integrating theoretical derivations with practical implementation details. In this work, we summarize key components for comprehensive 1D blood flow simulations, including the derivation of reduced-order governing equations, the method of characteristics (Riemann invariants), a finite volume-based numerical scheme, boundary conditions (application of Riemann invariants for reflective/non-reflective and 3-element Windkessel outlet boundaries), junction treatments, verification of presented methodologies, and relevant practical applications. Additionally, we provide detailed step-by-step instructions for implementing the numerical scheme, applying boundary conditions, and treatment of junctions. By integrating rigorous theory with practical guidance for implementation, we seek to improve accessibility of 1D blood flow simulations. We anticipate that this guide will serve as a valuable resource and foundational reference for both novice and experienced researchers in cardiovascular modeling.
}




\maketitle

\section{Introduction}\label{Intro}
One-dimensional (1D) blood flow simulations are widely used in biomedical applications due to their computational efficiency compared to full three-dimensional (3D) models. Although 1D simulations offer less spatial detail than 3D simulations, they can capture pulse wave dynamics (area, pressure, and volume flow rate) and systemic responses in complex geometries that would be prohibitively expensive in 3D. Computational speed allows the 1D models to be particularly valuable for parameter sensitivity analyses and large-scale vascular network analyses. Over the past few decades, 1D models have been used 
in various contexts, including analyzing waveform propagation through the major human arterial network~\cite{wang20141d, sherwin2003computational, epstein2015reducing} and circle of Willis~\cite{huang20181d, alastruey2007modelling}, as well as for investigating cardiovascular diseases such as atherosclerosis~\cite{el2019mathematical}, hypertension~\cite{pfaller2022automated}, and stenosis~\cite{kozitza2024estimating, ghigo2018time}. One-dimensional modeling is versatile, enabling integration with 3D simulations, which has previously been employed to assess the influence of inflow boundary conditions from the heart to the carotid artery~\cite{blanco2009potentialities, blanco2010assessing} as well as to examine the impact of aneurysms~\cite{ blanco2009potentialities}.

Although open-source codes for 1D blood flow simulations are available~\cite{benemerito2024openbf, diem2017vampy, updegrove2017simvascular}, they typically lack a unified, step-by-step guide that integrates both theoretical derivations and practical details for implementation. In particular, new users seeking to implement and adapt 1D models would benefit from a concise and clear guide documenting common boundary conditions and numerical schemes. This article is intended to provide such a resource, which may prove valuable to individuals seeking to employ 1D modeling approaches for research, educational purposes, and/or biomedical applications. We intend for this manuscript to complement many other valuable resources summarizing various individual aspects of 1D modeling, including general overview~\cite{shi2011review}, mathematical derivation~\cite{formaggia2010cardiovascular}, numerical schemes~\cite{wang2015verification}, boundary conditions~\cite{cousins2012boundary, alastruey2008lumped}, and applications~\cite{alastruey2007modelling, formaggia2010cardiovascular, boileau2015benchmark}.

Here, we provide an overview of 1D blood flow modeling, with particular emphasis on implementation details. We first introduce the governing equations and their derivation, followed by an explanation of the method of characteristics. We then explain numerical schemes for solving these governing equations and discuss boundary condition treatments (inlet boundary, non-reflective outlet boundary, and three-element Windkessel model outlet boundary) using the method of characteristics. We also cover the treatment of junctions, including bifurcations and anastomoses. The presented methods are validated against a published benchmark study that compares different 1D numerical schemes with 3D simulations~\cite{boileau2015benchmark}. Finally, we illustrate how 1D blood flow modeling can be used for systemic arterial network predictions.

\newpage
\section{Symbols and Definitions}
\FloatBarrier
\begin{table}[h!]
\caption{List of symbols commonly used in 1D blood flow simulations and their definitions.}
  \centering
    \large
  \begin{tabular*}{0.8\textwidth}{@{\extracolsep{\fill}} >{\centering\arraybackslash}p{0.2\textwidth} >{\centering\arraybackslash}p{0.55\textwidth}}
    \hline
    \textbf{Symbol} & \textbf{Definition} \\
    \hline
    $A$ & Cross-sectional area of the vessel \\
    $P$ & Internal pressure of the vessel \\
    $Q$ & Volume flow rate through the vessel \\
    $u$ & Average axial velocity across the cross-section \\
    $\rho$ & Density of blood \\
    $\alpha$ & Momentum correction coefficient \\
    $\mu$ & Dynamic viscosity of blood \\
    $\gamma$ & Velocity profile correction factor \\
    $P_{ext}$ & External pressure on the vessel \\
    $h$ & Vessel wall thickness \\
    $E$ & Elastic modulus of the vessel \\
    $\nu$ & Poisson's ratio of the vessel wall \\
    $A_0$ & Initial area of the vessel at equilibrium \\
    $\lambda$ & Eigenvalue \\
    $W$ & Riemann invariants \\
    $R_1$ & Characteristic resistance \\
    $R_2$ & Resistance of the periphery \\
    $C$ & Capacitance of the periphery \\
    \hline
  \end{tabular*}
  \label{tab:notation}
\end{table}

\section{Mathematical Background}
\begin{figure}[t!]
  \begin{center}
  \includegraphics[width=0.8\linewidth]{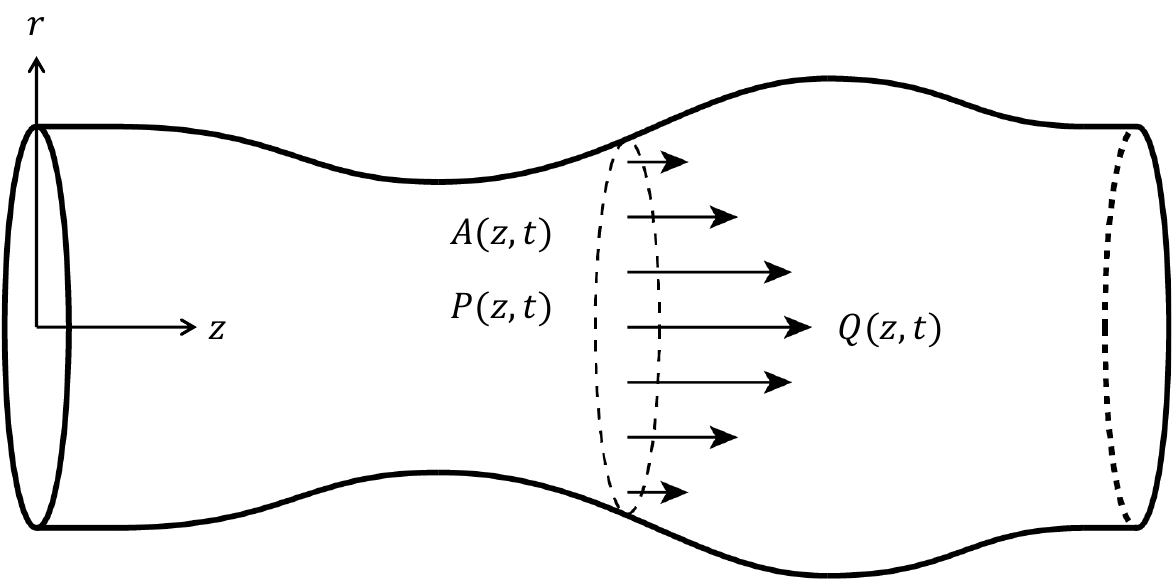}
  \end{center}
  \caption{Segment of a single compliant artery with its notations. $A$ is the cross-sectional area, $P$ is the internal pressure, and $Q$ is the volume flow rate through the vessel.} 
  \label{fig_exp:single blood vessel segment}
\end{figure}

Figure~\ref{fig_exp:single blood vessel segment} illustrates fluid flow through a compliant vessel that represents arterial flow along the axial direction $z$. The variable $A$ represents the vessel's cross-sectional area, $P$ denotes the internal pressure, and $Q$ is the volumetric flow rate through the vessel. For typical 1D blood flow simulations, blood is assumed to be incompressible and Newtonian, ensuring that both the blood density, $\rho$, and the dynamic viscosity, $\mu$, remain constant. 
The flow is characterized by three equations: conservation of mass, conservation of momentum, and a constitutive equation that relates $A$ and $P$. A detailed derivation of these 1D equations is provided below. 

\subsection{Governing Equations}
\subsubsection{Conservation of Mass}
The mass conservation (continuity) equation for fluid flow is expressed as: 
\begin{equation*}
    \frac{\partial \rho}{\partial t}+{\nabla} \cdot (\rho\mathbf{u})=0,
\end{equation*}
where $\rho$ is the fluid density, $t$ is time, and ${\bf u}$ is the velocity vector. Given that the fluid is incompressible, 
the density is constant and so the equation simplifies to:
\[\nabla \cdot \mathbf{u} = 0.\]
In cylindrical coordinates, ${\bf u}=\left (u_r, u_\theta, u_z \right )$ and assuming swirl-free ($v_{\theta}=0$) and axisymmetric flow ($\frac{\partial}{\partial \theta}=0$), the continuity equation becomes: 
\begin{equation}
\label{eqn:cylindrical_continuity}
\frac{\partial v_z}{\partial z} + \frac{1}{r}\frac{\partial}{\partial r}(rv_r) = 0.
\end{equation}
To obtain the 1D continuity equation, integrate equation~\eqref{eqn:cylindrical_continuity} over the cross-sectional area by multiplying the equation by $2\pi r$ and integrating with respect to $r$ from 0 to $R$:
\[2\pi \int_{0}^{R} \left(\frac{\partial v_z}{\partial z}r + \frac{\partial}{\partial r}(rv_r)\right) \,dr=0.\]
By applying Leibniz's rule for differentiation under the integral sign, one obtains
\begin{equation}
\label{eqn:cyl_continuity}
2\pi \frac{\partial}{\partial z}\int_{0}^{R} v_z r \, dr -2\pi \frac{\partial R}{\partial z}[rv_z]_{r=R} + 2\pi[rv_r]_{r=R} = 0.
\end{equation}
The volume flow rate, $Q$, is defined as 
\[Q = 2\pi\int_{0}^{R} v_z r dr.\]
Thus the first term in the equation \eqref{eqn:cyl_continuity} becomes $\frac{\partial Q}{\partial z}$.
Applying the no-slip condition at the wall, $v_z(r=R)=0$, causes the second term on the left-hand side to vanish. Furthermore, assuming the fluid velocity matches the vessel wall velocity, we have $v_r(r=R)=\frac{\partial R}{\partial t}$. Introducing the cross-sectional area $A = \pi R^2$, the third term simplifies to
\[2\pi[rv_r]_{r=R} = 2\pi R \frac{\partial R}{\partial t} = \frac{\partial A}{\partial t}.\]
Substituting these expressions into the continuity equation~\eqref{eqn:cylindrical_continuity}, we obtain
\begin{equation}
\label{eqn:1Dmass}
    \frac{\partial A}{\partial t} +\frac{\partial Q}{\partial z} = 0.
\end{equation}

\subsubsection{Conservation of Momentum}
The momentum conservation equation, commonly referred to as the Navier-Stokes equation, in the radial direction of a cylindrical coordinates system with swirl-free $(v_{\theta}=0)$ and axisymmetric $(\sfrac{\partial}{\partial \theta}=0)$ assumptions is expressed as:
\[\rho \left ( \frac{\partial v_r}{\partial t} + v_r\frac{\partial v_r}{\partial r} + v_r\frac{\partial v_r}{\partial z} \right ) = -\frac{\partial P}{\partial r} + \mu \left[\frac{1}{r}\frac{\partial}{\partial r}\left( r \frac{\partial v_r}{\partial r}\right) + \frac{\partial^2 v_r}{\partial z^2} - \frac{v_r}{r^2} \right].\]
After applying the long-wave approximation~\cite{wang20141d} (the axial length is much greater the radial length) and performing the appropriate scaling, the above equation reduces to 
\[\frac{1}{\rho}\frac{\partial P}{\partial r} = 0.\]
This indicates that the pressure remains constant in the radial direction. Building on the previously stated assumptions, the momentum equation in the axial direction is formulated as follows: 
\begin{equation}
\label{eqn:cylindrical_momentum}
    \rho \left(\frac{\partial v_z}{\partial t} + v_r\frac{\partial v_z}{\partial r} + v_z\frac{\partial v_z}{\partial z} \right) = -\frac{\partial P}{\partial z} + \frac{\mu}{r}\frac{\partial}{\partial r}\left( r\frac{\partial v_z}{\partial r}\right).
\end{equation}
To obtain the 1D momentum equation, integrate equation~\eqref{eqn:cylindrical_momentum} over the cross-sectional area by multiplying the equation by $2\pi r$ and integrating with respect to $r$ from 0 to $R$:
\begin{equation}
\label{eqn:integral1}
    2\pi\left[\int_{0}^{R} r\frac{\partial v_z}{\partial t}\,dr + \int_{0}^{R} \left(rv_r\frac{\partial v_z}{\partial r} + rv_z \frac{\partial v_z}{\partial z}\right)\,dr + \int_{0}^{R} \frac{r}{\rho} \frac{\partial P}{\partial z}\,dr\right] = 2\pi\left[\int_{0}^{R} \frac{\mu}{\rho} \frac{\partial}{\partial r}\left(r\frac{\partial v_z}{\partial r}\right)\,dr\right].
\end{equation}
The first term on the left-hand side after the Leibniz's rule represents the temporal change of momentum within a control volume:
\[\frac{\partial}{\partial t}\int_{0}^{R} 2\pi r v_z\, dr = \frac{\partial Q}{\partial t}.\]
The second term on the left-hand side (advection term), can be decomposed as follows by applying integration by parts to the first term:
\begin{equation}
    \label{eqn:integral2}
    2\pi\int_{0}^{R} \left(rv_r\frac{\partial v_z}{\partial r} + rv_z \frac{\partial v_z}{\partial z}\right)\,dr = 2\pi\int_{0}^{R} \frac{\partial(rv_r v_z)}{\partial r}\,dr - 2\pi\int_{0}^{R} v_z\frac{\partial(rv_r)}{\partial r}\,dr + 2\pi\int_{0}^{R} rv_z\frac{\partial v_z}{\partial z}\,dr.
\end{equation}

The first term on the right-hand side of equation~\eqref{eqn:integral2} can be evaluated as:
\[2\pi \int_0^R \frac{\partial(r v_r v_z)}{\partial r} dr=2\pi\left(\left[ r v_r v_z\right]_{r=R} -\left[rv_rv_z\right]_{r=0}\right).\]
Since $v_z(r=R) = 0$, the whole term vanishes. 
Utilizing the continuity equation, we have $\frac{\partial (rv_r)}{\partial r} = -r\frac{\partial v_z}{\partial z}$. Applying this relation to the second term on the right-hand side of equation~\eqref{eqn:integral2} simplifies the expression to:
\[2\pi\int_{0}^{R} \left(rv_r\frac{\partial v_z}{\partial r} + rv_z \frac{\partial v_z}{\partial z}\right)\,dr = 4\pi\int_{0}^{R} rv_z\frac{\partial v_z}{\partial z}\,dr.\]
Noting that $v_z\frac{\partial v_z}{\partial z} = \frac{1}{2}\frac{\partial(v_z)^2}{\partial z}$ by the chain rule, the 
right-hand side can be simplified as:
\[2\pi\int_{0}^{R} \left(rv_r\frac{\partial v_z}{\partial r} + rv_z \frac{\partial v_z}{\partial z}\right)\,dr = 2\pi\frac{\partial}{\partial z}\int_{0}^{R} r {v_z}^2\,dr.\]
We define the integral of $rv_z^2$ as:
\[2\pi\int_{0}^{R} r{v_z}^2 = \alpha\frac{Q^2}{A}.\]
Here, $\alpha$, known as the momentum correction coefficient or Coriolis coefficient, is defined as:
\[\alpha = \frac{2\pi A}{Q^2}\int_{0}^{R} r {v_z}^2 \,dr = \frac{2\pi\int_{0}^{R}2\pi r\,dr}{\left(\int_{0}^{R} r v_z \,dr\right)^2}\int_{0}^{R} r {v_z}^2 \,dr.\]
The value of $\alpha$ is determined by the axial velocity profile~\cite{formaggia2003one, wang2015verification}. The Womersley number, defined as $Wo = R\sqrt{\frac{\rho \omega}{\mu}}$ (with $\omega$ representing the angular frequency of the pulsatile flow), plays a crucial role in determining the shape of the axial velocity profile. For high Womersley number ($Wo >> 1$), $\alpha=1$ is appropriate in accordance with experiments, whereas for low Womersley number ($Wo << 1$), $\alpha=\frac{4}{3}$ is used. 

\begin{figure}[t!]
    \centering
    \includegraphics[width=0.6\linewidth]{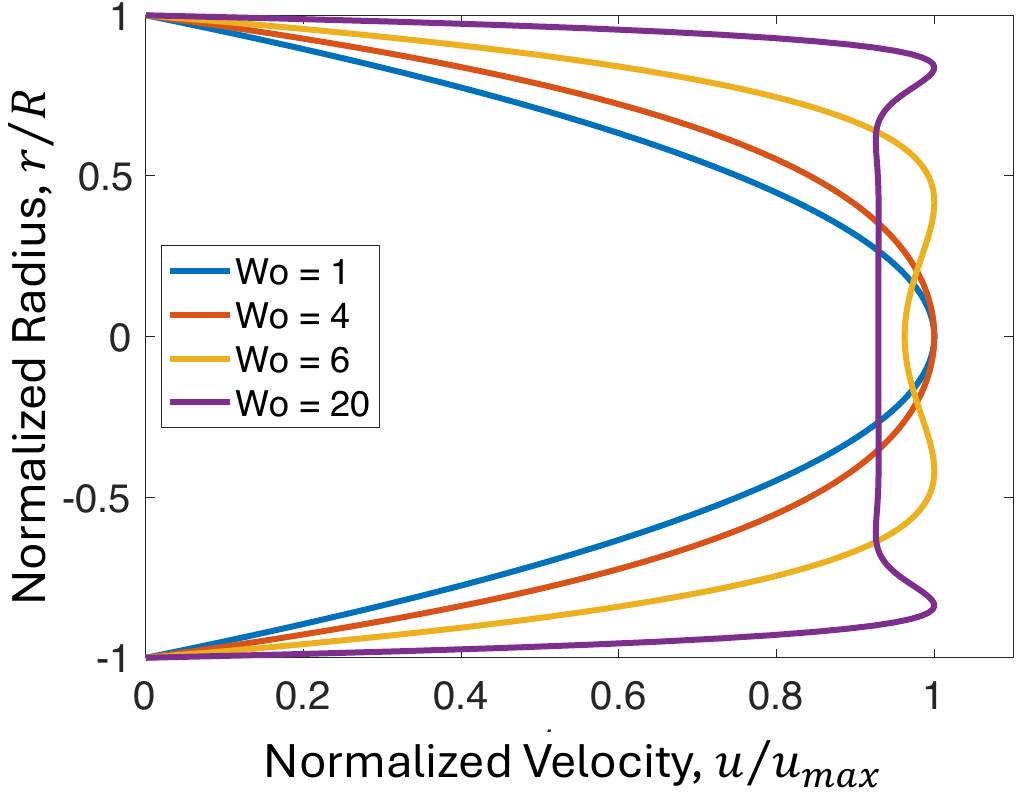}
    \caption{Representative instantaneous velocity profiles at maximum velocity for various Womersley numbers. Higher Womersley numbers yield a flatter velocity profile.}
    \label{fig_exp: velocity profiles vs Wo}
\end{figure}

The integration of the third term on the left-hand side of equation~\eqref{eqn:integral1} is straightforward, noting that $\sfrac{\partial P}{\partial r}=0$, 
\[2\pi\int_{0}^{R}\frac{r}{\rho}\frac{\partial P}{\partial z}\,dr = \frac{A}{\rho}\frac{\partial P}{\partial z}.\]
Finally, the right-hand side (viscous) term, can be integrated as follows:
\[2\pi\int_{0}^{R}\frac{\mu}{\rho}\frac{\partial}{\partial r}\left(r\frac{\partial v_z}{\partial r}\right)\,dr = 2\pi\frac{\mu}{\rho}\left[r\frac{\partial v_z}{\partial r}\right]_{r=R}.\]
For the viscous term, it is often assumed that the flow is parabolic (Poiseuille flow). With this assumption, the wall shear stress is proportional to the volumetric flow rate as follows:
\[\tau_w = \mu\left.\frac{\partial v_z}{\partial r}\right|_{r=R} = -\frac{4\mu Q}{\pi R^3}.\]
By substituting this expression into the previous equation, we obtain 
\[2\pi\int_{0}^{R}\frac{\mu}{\rho}\frac{\partial}{\partial r}\left(r\frac{\partial v_z}{\partial r}\right)\,dr = -K_R \frac{Q}{A}\]
where $K_R = 8\pi\frac{\mu}{\rho}$.

A more general form of $K_R$, not restricted to a parabolic velocity profile, can be derived~\cite{hughes1973one}. For an arbitrary axisymmetric flow with the no-slip boundary condition imposed at $r=R$, one obtains: 
\[v_z = U\frac{\gamma+2}{\gamma}\left[1-\left(\frac{r}{R}\right)^\gamma\right],\]
where $U$ represents the maximum velocity at the center of the cross-section and $\gamma$ is the profile coefficient.
Since $\gamma$ is a constant, the viscous term ($2\pi\frac{\mu}{\rho}\left[r\frac{\partial v_z}{\partial r}\right]_{r=R}$) can be written as $-2\pi\frac{\mu}{\rho}(\gamma+2)U$. Although the momentum correction coefficient $\alpha$ is fixed at 1 for most cases, the velocity profile coefficient $\gamma$ varies depending on $Wo$. For the parabolic flow as introduced above, $\gamma = 2$ is often assumed leading to $K_R = 8\pi\frac{\mu}{\rho}$, which is widely used. For the flatter profile, $\gamma = 9$ is used and it agrees well with experimental data~\cite{smith2002anatomically}, leading to $K_R = 22\pi\frac{\mu}{\rho}$. Finally, equation~\eqref{eqn:cylindrical_momentum} can be written in the form:
\begin{equation}
\label{eqn:1Dmomentum}
    \frac{\partial Q}{\partial t} + \frac{\partial}{\partial z}\left(\alpha\frac{Q^2}{A}\right) + \frac{A}{\rho}\frac{\partial P}{\partial z} = -K_R \frac{Q}{A}.
\end{equation}

In summary, the governing equations for 1D blood flow, with the assumption of momentum correction coefficient $\alpha = 1$ (i.e., large $Wo$), can be written as:
\begin{equation}
\label{eqn:1Dgoverning_equations}
\begin{aligned}
    \frac{\partial A}{\partial t} + \frac{\partial Q}{\partial z} &= 0\\ 
    \frac{\partial Q}{\partial t} + \frac{\partial}{\partial z}\left(\frac{Q^2}{A}\right) 
    + \frac{A}{\rho}\frac{\partial P}{\partial z} &= -K_R \frac{Q}{A}.
\end{aligned}
\end{equation}
Alternatively, the governing equations can be formulated in terms of the cross-sectional area $A$ and the axial velocity $u$. In this formulation, the volumetric flow rate $Q$ naturally follows the relation $Q=A\;u$. Using $(A,u)$ variable system, the equation \eqref{eqn:1Dmass} can be written as:
\[\frac{\partial A}{\partial t} + \frac{\partial (Au)}{\partial z} = 0.\]
The equation \eqref{eqn:1Dmomentum} can be written as:
\[\frac{\partial(Au)}{\partial t} + \frac{\partial}{\partial z}\left( Au^2\right) + \frac{A}{\rho}\frac{\partial P}{\partial z} = -K_R\;u.\]
By applying the chain rule, the 1D continuity (conservation of mass) equation becomes:
\[\frac{\partial A}{\partial t} + A\frac{\partial u }{\partial z} + u \frac{\partial A}{\partial z} = 0. \]
Similarly, the 1D Navier-Stokes (conservation of momentum) equation can be written as:
\[A \frac{\partial u}{\partial t}+u\frac{\partial A}{\partial t} + 2Au\frac{\partial u}{\partial z} + u^2 \frac{\partial A}{\partial z} + \frac{A}{\rho}\frac{\partial P}{\partial z}= -K_R u.\]
Substituting $\frac{\partial A}{\partial t} =- A\frac{\partial u }{\partial z} - u \frac{\partial A}{\partial z}$ from the continuity equation into the 1D Navier-Stokes equation leads to: 
\[A\frac{\partial u}{\partial t}+u\left(-A\frac{\partial u}{\partial z} - u\frac{\partial A}{\partial z} \right) + 2Au \frac{\partial u}{\partial z} + u^2 \frac{\partial A}{\partial z} + \frac{A}{\rho}\frac{\partial P}{\partial z} = -K_R u.\]
After further manipulation, the system of equations can be equivalently expressed using $A$ and $u$:
\begin{equation}
\label{eqn:1Dgoverning_equations_Au}
\begin{aligned}
    \frac{\partial A}{\partial t} + \frac{\partial Au}{\partial z} &= 0 \\ 
    \frac{\partial u}{\partial t} + u \frac{\partial u}{\partial z} + \frac{1}{\rho}\frac{\partial P}{\partial z} &= -K_R\frac{u}{A}.
\end{aligned}
\end{equation}

\subsection{Constitutive equation}
The governing equations~\eqref{eqn:1Dmass} and \eqref{eqn:1Dmomentum} have three variables: $A$, $Q$, and $P$. Since we have two equations and three unknowns, an additional equation is needed to close the system. This is typically accomplished using a nonlinear constitutive equation that relates $A$ and $P$~\cite{formaggia2003one, olufsen2000numerical, mynard20081d}:
\begin{equation}
    \label{eqn:constitutive pressure}
    P = P_{ext} + \beta(\sqrt{A}-\sqrt{A_0})
\end{equation}
where $P_{ext}$ denotes the external pressure from the surrounding tissue, $A_0$ is the initial area at the equilibrium state (i.e., when $P=P_{ext}$), and $\beta$ is related to the vessel's mechanical properties and is defined as
\[\beta = \frac{\sqrt{\pi}hE}{A_0 (1-\nu^2)},\]
where $h$ is the vessel wall thickness, $E$ is the Young's modulus, and $\nu$ is the Poisson ratio. The value $\nu=0.5$ is most widely used in the literature since the vessel tissue is considered incompressible. Alternative relationships linking $A$ and $P$ for linear elastic, nonlinear elastic, collapsible tube, and viscoelastic models are detailed in ref.~\cite{mynard2007one}. 

By differentiating the constitutive pressure equation~\eqref{eqn:constitutive pressure} with respect to $z$ using the chain rule, the pressure term can be incorporated into the momentum equation~\eqref{eqn:1Dmomentum} as follows:
\[\frac{\partial P}{\partial z} = \frac{\partial P_{ext}}{\partial z} + \frac{\partial \beta}{\partial z}(\sqrt{A}-\sqrt{A_0}) + \frac{\beta}{2\sqrt{A}}\frac{\partial A}{\partial z} - \frac{\beta}{2\sqrt{A_0}}\frac{\partial A_0}{\partial z}.\]
It is commonly assumed that the material properties $A_0$ and $\beta$ are uniform along the length of a vessel segment, implying
\[\frac{\partial A_0}{\partial z}=0 \quad \text{and} \quad \frac{\partial \beta}{\partial z}=0.\]
Additionally, by often setting the external pressure to zero ($P_{ext}=0$), the corresponding terms are eliminated from the momentum equation, leading to a simplified formulation:
\begin{equation}
\label{eqn: dpdz term}
    \frac{\partial P}{\partial z} = \frac{\beta}{2\sqrt{A}}\frac{\partial A}{\partial z}.
\end{equation}

\section{Characteristics of equation}

\subsection{Conservative form}
The system of governing equations~\eqref{eqn:1Dgoverning_equations_Au} can be written in conservative form as 
\begin{equation}
\label{eqn:conservative form}
    \frac{\partial U}{\partial t} + \frac{\partial F(U)}{\partial z} = S(U),
\end{equation}
where 
\[ U = \begin{bmatrix} A \\ u \end{bmatrix},\qquad F = \begin{bmatrix}
    Au \\ \frac{u^2}{2} + \frac{P}{\rho}
\end{bmatrix},\qquad S = \begin{bmatrix}
    0 \\ -K_R\frac{u}{A}
\end{bmatrix}.\]
Here, $U$ is the vector of conservative variables, $F(U)$ is the flux vector, and $S(U)$ represents the source terms. 

Since $\frac{1}{\rho}\frac{\partial P}{\partial A} = \frac{\beta}{2\rho \sqrt{A}}$ (obtained by taking the derivative of equation~\eqref{eqn:constitutive pressure}), the flux Jacobian matrix can be computed as 
\begin{equation}
\label{eqn: Jacobian matrix}
    J(U) = \frac{\partial F}{\partial U} = \begin{bmatrix}
    \sfrac{\partial F_1}{\partial U_1} & \sfrac{\partial F_1}{\partial U_2} \\ \sfrac{\partial F_2}{\partial U_1} & \sfrac{\partial F_2}{\partial U_2}
\end{bmatrix} = \begin{bmatrix}
    u & A \\ \frac{\beta}{2\rho \sqrt{A}} & u
\end{bmatrix},
\end{equation}
where the subscripts ${1,2}$ index the entries of $F$ and $U$. These formulations are crucial for the discussion in the following chapter.

\subsection{Characteristic variables}
The eigenvalues $\lambda_{1,2}$ of the Jacobian matrix \eqref{eqn: Jacobian matrix} are determined by solving
\[det(H - \lambda I) = 0.\]
Substituting the Jacobian matrix $J$ into this equation yields
\[\begin{vmatrix}
    u - \lambda & A \\
    \frac{\beta}{2\rho \sqrt{A}} & u - \lambda
\end{vmatrix} = 0\]
which has eigenvalues 
\[\lambda_{1,2} = u \pm c,\]
or in the $A$ and $Q$ system,
\[\lambda_{1,2} = \frac{Q}{A} \pm c.\]
Here, $c = \sqrt{\frac{\beta}{2\rho}\sqrt{A}}$ is known as the Moens-Korteweg celerity, which represents the speed at which small perturbations propagate along the vessel. Physiologically, $A > 0$, $\rho > 0$, and $\beta > 0$, which ensures that the eigenvalues are real and distinct. This confirms that the system of equations \eqref{eqn:1Dgoverning_equations_Au} is strictly hyperbolic. Additionally, under typical arterial conditions, the eigenvalues satisfy $\lambda_1 = u+c > 0$ and $\lambda_2 = u-c < 0$, indicating that the system is subsonic~\cite{formaggia2010cardiovascular}.

The left eigenvectors, $L = \begin{bmatrix}
    l_1 & l_2
\end{bmatrix}$ are obtained by solving 
\begin{equation}
    \label{eqn:left eigenvectors}
    LJ = \Lambda L,
\end{equation}
where $H$ is the flux Jacobian and $\Lambda$ is the diagonal matrix of eigenvalues. After appropriate scaling, the left eigenvector matrix can be expressed as 
\[L = \begin{bmatrix}
    l_1^T \\ l_2^T 
\end{bmatrix} = \begin{bmatrix}
    \frac{c}{A} & 1 \\ 
    -\frac{c}{A} & 1
\end{bmatrix}.\]
Equation \eqref{eqn:left eigenvectors} can be rewritten as $J = L^{-1}\Lambda L$. Substituting this expression for $J$ (noting that $\frac{\partial F}{\partial z} = \frac{\partial F}{\partial U}\frac{\partial U}{\partial z} = J\frac{\partial U}{\partial z}$) into the quasi-linear form (the highest-order derivative terms appear linearly) of the governing equations, we obtain
\[\frac{\partial U}{\partial t} + L^{-1}\Lambda L \frac{\partial U}{\partial z} = f.\]
If we assume that the friction term $f$ is negligible near the boundaries~\cite{mynard20081d, formaggia2003one}, the equation simplifies to 
\[L\frac{\partial U}{\partial t} + \Lambda L \frac{\partial U}{\partial z} = 0.\]
By introducing $\partial W = L \partial U$, the characteristic system becomes 
\begin{equation}
\label{eqn:Riemann invariants equations}
\begin{aligned}
\frac{\partial W}{\partial t} + \Lambda \frac{\partial W}{\partial z} = 0,
\end{aligned}
\end{equation}
which can be written in general form as 
\begin{align*}
    \frac{\partial W_1}{\partial t} + \lambda_1 \frac{\partial W_1}{\partial z} &= 0\\
    \frac{\partial W_2}{\partial t} + \lambda_2 \frac{\partial W_2}{\partial z} &= 0.
\end{align*}
Here, $W_1$ and $W_2$ are the characteristic variables (or Riemann invariants) of the system. They are derived by solving the relation $\frac{\partial W}{\partial U} = L$. They can be computed as
\begin{equation}
    \label{eqn:Riemann invariants}
    \begin{aligned}
        W_1 = u+4c \\ 
        W_2 = u-4c.
    \end{aligned}
\end{equation}
Since $W_1$ and $W_2$ are constant along the characteristic curves defined by $\frac{dz}{dt} = \lambda_{1,2}$, they serve as natural invariants of the flow. 

The original variables $(A, u)$ can be computed from the Riemann invariants via the relations 
\begin{equation}
    \label{eqn:Riemann invariants - in A and u}
    \begin{aligned}
        A &= \frac{(W_1-W_2)^4}{1024} \left(\frac{\rho}{\beta}\right)^2  \\ 
        u &= \frac{W_1+W_2}{2},
    \end{aligned}
\end{equation}
or in $(A, Q)$ system:
\begin{equation}
    \label{eqn:Riemann invariants - in A and Q}
    \begin{aligned}
        A &= \frac{(W_1-W_2)^4}{1024} \left(\frac{\rho}{\beta}\right)^2  \\ 
        Q &= A \frac{W_1+W_2}{2}.
    \end{aligned}
\end{equation}
Before delving deeper into how the Riemann invariants are used to handle boundary conditions, it is important to introduce the method of characteristics.

\subsection{Method of Characteristics}
Consider a first-order homogeneous hyperbolic partial differential equation (a wave equation) in quasi-linear form:
\[\frac{\partial v}{\partial t}+a \frac{\partial v}{\partial x}=0, \qquad v(x,t=0) = f(x).\]
Note that this is provided as an illustrative example; the variables $v$ and $a$ here are used for demonstration purposes.

\begin{figure}[t!]
    \centering
    \includegraphics[width=\linewidth]{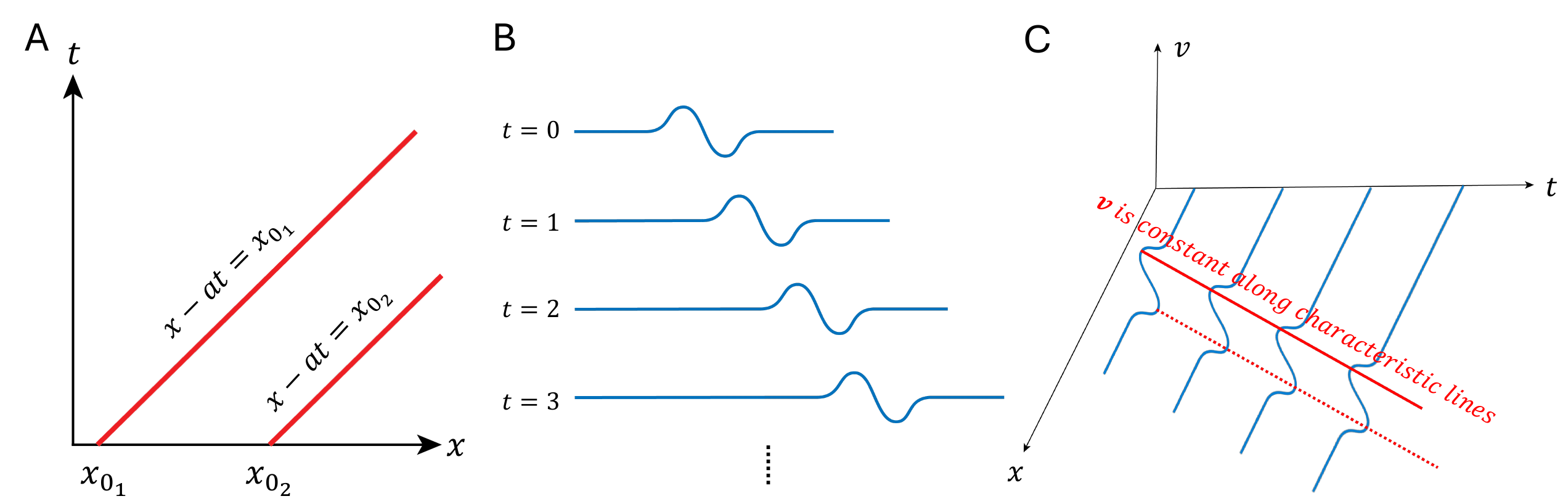}
    \caption{Representative examples of characteristic curves. (A) A set of characteristic curves. Each line is determined by its initial conditions, $x_{0_1}$ and $x_{0_2}$. (B) Demonstration of propagation of a wave in space (horizontal-axis) and time (vertical-axis). (C) Demonstration of characteristic lines that shows variable $v$ is conserved along those lines.}
    \label{fig_exp: method of characteristics 2}
\end{figure}

Instead of finding the solution $v(x,t)$ through separation of variables or numerical methods, we can use the method of characteristics by analyzing the characteristic curves along which $v$ remains constant. By applying the chain rule, we have
\[\frac{d}{dt}\left [ v(x(t),t)\right ] = \frac{dv}{dt}= \frac{dv}{dx} \frac{dx}{dt}+\frac{dv}{dt}.\]
If we assume that the characteristic curves satisfy $\frac{dx}{dt} = a$, and that $\frac{dv}{dt}=0$, this implies that $v$ does not change along these curves. The chain rule expression simplifies to 
\[\frac{dv}{dt}+a\frac{dv}{dx}=0.\]
Solving the ordinary differential equation $\frac{dx}{dt}=a$, we obtain
\[x = a t+x_0,\]
where $x_0$ is the initial position. Since the initial condition is given by $v(x,0)=f(x)$, the below equation can be derived along the characteristic curve of $x-at=x_0$:
\[v(x,t) = f(x_0) = f(x-at).\]
This means that the solution for $\frac{dv}{dt}+a\frac{dv}{dx}=0$ is $v(x,t)=f(x-at)$. In summary, the method of characteristics works as follows: to determine the value of $v$ at a point $(x,t)$, trace the characteristic curve that passes through $(x,t)$ back to where it intersects the initial condition. The value of $v$ at $(x,t)$ is then given by the initial condition at that intersection point. Figure \ref{fig_exp: method of characteristics 2} helps to visualize this concept. 
This concept is applied to equation \eqref{eqn:Riemann invariants equations}, particularly for determining boundary conditions. More details on this application is presented in the Boundary Conditions (\ref{sec: Boundary conditions}) and Conjunctions (\ref{sec: Conjunctions}) sections.

\section{Numerical Scheme}
There are many numerical schemes available for solving hyperbolic partial differential equations (PDEs), particularly the 1D blood flow equations \eqref{eqn:conservative form}. These methods include finite difference schemes (\cite{olufsen2000numerical, reymond2009validation, saito2011one}), finite volume schemes (\cite{delestre2013well, wang2015verification, melis2017gaussian}), and finite element methods (\cite{formaggia2003one, mynard2007one, grinberg2011modeling, alastruey2011pulse, boileau2015benchmark}). Among these, the finite element methods, especially the discontinuous Galerkin method~\cite{reed1973triangular}, are particularly well-known. In this work, however, we introduce a finite volume scheme based on the Monotonic Upstream-centered Scheme for Conservation Laws (MUSCL) due to its relative simplicity in implementation and numerical accuracy. The explanation presented below concisely summarizes the approaches outlined in Delestre et al. \cite{delestre2013well} and Wang et al. \cite{wang2015verification}.

\subsection{Finite Volume - MUSCL scheme}
The governing equation in the conservative form $(A,Q)$ system \eqref{eqn:conservative form} is given by
\[\frac{\partial U}{\partial t} + \frac{\partial F}{\partial z} = S\]
where 
\[ U = \begin{bmatrix} A \\ Q \end{bmatrix},\qquad F = \begin{bmatrix}
    Q \\ \frac{Q^2}{A} + \frac{\beta}{3\rho} A^{\sfrac{3}{2}}
\end{bmatrix},\qquad S = \begin{bmatrix}
    0 \\ -K_R\frac{Q}{A}
\end{bmatrix}.\] This formulation is obtained by combining equations \eqref{eqn:1Dgoverning_equations} and  \eqref{eqn: dpdz term}.
In the finite volume method, the spatial domain is subdivided into cells (or finite volumes). For a cell with center $i$ and boundaries at $\left[z_{i-\sfrac{1}{2}}, \enspace z_{i+\sfrac{1}{2}}\right]$, we integrate the governing equation over the cell:
\[
\int_{z_{i-\sfrac{1}{2}}}^{z_{i+\sfrac{1}{2}}} \frac{\partial U}{\partial t} dz + \int_{z_{i-\sfrac{1}{2}}}^{z_{i+\sfrac{1}{2}}} \frac{\partial F}{\partial z} dz = \int_{z_{i-\sfrac{1}{2}}}^{z_{i+\sfrac{1}{2}}} S dz.
\]
Applying the divergence theorem to the flux term $(F)$ yields
\[
\int_{z_{i-\sfrac{1}{2}}}^{z_{i+\sfrac{1}{2}}} \frac{\partial U}{\partial t} dz + F|_{z_{i+\sfrac{1}{2}}} - F|_{z_{i-\sfrac{1}{2}}} = \int_{z_{i-\sfrac{1}{2}}}^{z_{i+\sfrac{1}{2}}} S dz
.\]
For cell $i$, the volume-averaged value of $U$ and $S$ can be computed as
\[U_i = \frac{1}{z_{i+\sfrac{1}{2}}-z_{i-\sfrac{1}{2}}}\int_{z_{i-\sfrac{1}{2}}}^{z_{i+\sfrac{1}{2}}}U(z) dz, \quad S_i = \frac{1}{z_{i+\sfrac{1}{2}}-z_{i-\sfrac{1}{2}}}\int_{z_{i-\sfrac{1}{2}}}^{z_{i+\sfrac{1}{2}}}S(z) dz.\]
Since the cell width is $\Delta z=z_{i+\sfrac{1}{2}}-z_{i-\sfrac{1}{2}}$, the semi-discrete form of the governing equation becomes
\begin{equation}
    \label{eqn:MUSCL_governing equation}
    \frac{d U_i}{dt} = -\frac{F|_{z+\sfrac{1}{2}}-F|_{z-\sfrac{1}{2}}}{\Delta z} + S_i.
\end{equation}

We use the Rusanov flux, also called as the Lax-Friedrichs flux, to approximate the inter-cell fluxes~\cite{bouchut2004nonlinear}. At the cell interface located at $z={i\pm\sfrac{1}{2}}$, the numerical flux is computed as
\begin{align*}
    F|_{i+\sfrac{1}{2}} &= \frac{1}{2}\left(F(U_{i+\sfrac{1}{2}}^R)+F(U_{i+\sfrac{1}{2}}^L)\right) - \frac{c_{i+\sfrac{1}{2}}}{2}\left(U_{i+\sfrac{1}{2}}^R - U_{i+\sfrac{1}{2}}^L\right)
    \\ 
    F|_{i-\sfrac{1}{2}} &= \frac{1}{2}\left(F(U_{i-\sfrac{1}{2}}^R)+F(U_{i-\sfrac{1}{2}}^L)\right) - \frac{c_{i-\sfrac{1}{2}}}{2}\left(U_{i-\sfrac{1}{2}}^R - U_{i-\sfrac{1}{2}}^L\right),
\end{align*}
where $U^R$ and $U^L$ are defined further below, and $c$ is the maximum characteristic speed at the cell interface computed as
\begin{align*}
c_{i+\sfrac{1}{2}} &= max \left(\lambda_1 \left(U_{i+\sfrac{1}{2}}^R\right), \enspace \lambda_1 \left(U_{i+\sfrac{1}{2}}^L\right) \right), \\
c_{i-\sfrac{1}{2}} &= max \left(\lambda_1 \left(U_{i-\sfrac{1}{2}}^R\right), \enspace \lambda_1 \left(U_{i-\sfrac{1}{2}}^L\right) \right)
\end{align*}
with the eigenvalue $\lambda_1$ given by $\lambda_1 = \frac{Q}{A}+\sqrt{\frac{\beta}{2\rho}\sqrt{A}}$ as described previously. 

To limit spurious oscillations near discontinuities or shocks, it is necessary to reconstruct the left $(U^L)$ and right $(U^R)$ states at each cell interface. These interface values are computed using a slope-limited reconstruction:
\begin{align*}
U_{i+\sfrac{1}{2}}^R = U_{i+1} - \frac{\Delta z}{2} DU_{i+1}, \qquad U_{i+\sfrac{1}{2}}^L = U_{i} + \frac{\Delta z}{2} DU_{i}, \\
U_{i-\sfrac{1}{2}}^R = U_{i} - \frac{\Delta z}{2} DU_{i}, \qquad U_{i-\sfrac{1}{2}}^L = U_{i-1} + \frac{\Delta z}{2} DU_{i-1}
\end{align*}
where $DU_i$ is the reconstructed sloped within cell $i$. 

\begin{figure}[t!]
    \centering
    \includegraphics[width=0.7\linewidth]{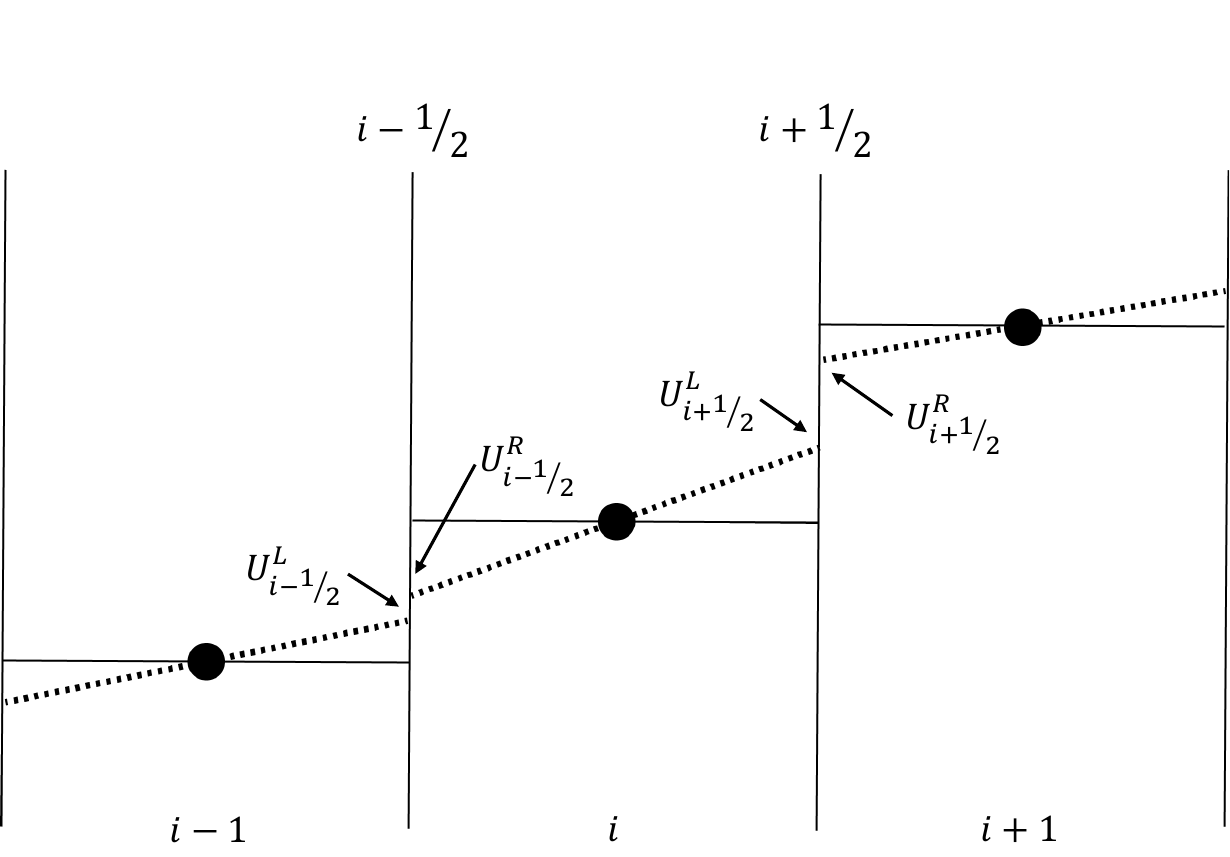}
    \caption{Representative example of left and right linear extrapolations of variables for MUSCL scheme.}
    \label{fig:MUSCL reconstruction}
\end{figure}

There are several approaches to reconstruct the variables using a slope limiter. In our implementation, we adopt the \textit{minmod} slope limiter, which is defined as follows:
\[
\text{minmod}(x,y) = 
\begin{cases}
\min(x,y) & \text{if } x, y \ge 0, \\
\max(x,y) & \text{if } x, y \le 0, \\
0 & \text{otherwise}.
\end{cases}
\]
With this definition, the slope $DU_i$ is given by 
\[DU_i = \text{minmod}\left(\frac{U_i-U_{i-1}}{\Delta z}, \enspace \frac{U_{i+1}-U_{i}}{\Delta z}\right).\]
Figure~\ref{fig:MUSCL reconstruction} illustrates a representative example of the left and right linear extrapolations of the variables used in the presented MUSCL scheme.

After the spatial discretization using the slopes and the corresponding slope limiter, the semi-discrete form of the governing equations becomes
\[ \frac{dU_i}{dt} = \Phi(U_{i-2},... U_{i+2}) \coloneq -\frac{F|_{z+\sfrac{1}{2}}-F|_{z-\sfrac{1}{2}}}{\Delta z} + S_i\]
where $\Phi$ is a spatial operator that uses a five-point stencil.

For the temporal integration, we apply a two-step, second-order Adams-Bashforth scheme:
\[U^{n+1} = U^n + \Delta t \left(\frac{3}{2}\Phi\left(U^n\right) - \frac{1}{2} \Phi \left(U^{n-1}\right) \right)\]
where $U^n$ represents the solution at the current time step and $U^{n-1}$ represents the solution at the previous time step. 

\subsection{Steps for MUSCL implementation}
It is important to note that this presented scheme is completely explicit and requires a five-point stencil $(i-2,\,i-1,\,i,\,i+1,\,i+2)$ to compute the spatial operator (specifically, the flux term). Consequently, one ghost cell is needed at each end of the computational domain for variable reconstructions. These ghost cells are assigned the same values as those in the adjacent boundary cells. In summary, the MUSCL scheme for 1D blood flow can be implemented as follows:

\begin{enumerate}
    \item \textbf{Set the ghost cells at the boundaries}\\
    Assign the ghost cells to have the same values as their neighboring boundary cells:
    \begin{equation*}
        \begin{aligned}
            A_0&=A_1,\quad Q_0=Q_1,\\
            A_{n+1}&=A_n, \quad Q_{n+1} = Q_n.
        \end{aligned}
    \end{equation*}
    \item \textbf{Reconstruct the variables $U$ using the slopes}\\
    The reconstructed values at the cell interfaces are obtained via the \textit{minmod} slope limiter:
    \begin{equation*}
    \begin{aligned}
    U_{i+\sfrac{1}{2}}^R &= U_{i+1} - \frac{\Delta z}{2}\text{minmod}\left(\frac{U_{i+1}-U_{i}}{\Delta z}, \frac{U_{i+2}-U_{i+1}}{\Delta z}\right)\\
    U_{i+\sfrac{1}{2}}^L &= U_i + \frac{\Delta z}{2}\text{minmod}\left(\frac{U_i-U_{i-1}}{\Delta z}, \frac{U_{i+1}-U_i}{\Delta z}\right)\\
    U_{i-\sfrac{1}{2}}^R &= U_i - \frac{\Delta z}{2}\text{minmod}\left(\frac{U_i-U_{i-1}}{\Delta z}, \frac{U_{i+1}-U_i}{\Delta z}\right)\\
    U_{i-\sfrac{1}{2}}^L &= U_{i-1} + \frac{\Delta z}{2}\text{minmod}\left(\frac{U_{i-1}-U_{i-2}}{\Delta z}, \frac{U_{i}-U_{i-1}}{\Delta z}\right).
    \end{aligned}
    \end{equation*}
    Recall that $U$ represents the variable matrix $(A, Q)$; thus, it can be substituted directly by $A$ and $Q$ for reconstructions.  
    \item \textbf{Compute the maximum characteristic speed $c$ using the eigenvalue $\lambda$} \\
    To compute the flux term eventually, the characteristic speed $c$ is first evaluated in the interior cells. For example, 
    \[c_{i+\sfrac{1}{2}} = max \left(\lambda_1 \left(U_{i+\sfrac{1}{2}}^R\right), \enspace \lambda_1 \left(U_{i+\sfrac{1}{2}}^L\right) \right)\]
    where 
    \[\lambda_1 \left(U_{i+\sfrac{1}{2}}^R\right)=\frac{Q_{i+\sfrac{1}{2}}^R}{A_{i+\sfrac{1}{2}}^R}+\sqrt{\frac{\beta}{2\rho}\sqrt{A_{i+\sfrac{1}{2}}^R}}, \quad \lambda_1 \left(U_{i+\sfrac{1}{2}}^L\right)=\frac{Q_{i+\sfrac{1}{2}}^L}{A_{i+\sfrac{1}{2}}^L}+\sqrt{\frac{\beta}{2\rho}\sqrt{A_{i+\sfrac{1}{2}}^L}}.\] 
    \item \textbf{Compute the Rusanov fluxes} 
    \begin{align*}
    F|_{i+\sfrac{1}{2}} &= \frac{1}{2}\left(F(U_{i+\sfrac{1}{2}}^R)+F(U_{i+\sfrac{1}{2}}^L)\right) - \frac{c_{i+\sfrac{1}{2}}}{2}\left(U_{i+\sfrac{1}{2}}^R - U_{i+\sfrac{1}{2}}^L\right), 
    \\ 
    F|_{i-\sfrac{1}{2}} &= \frac{1}{2}\left(F(U_{i-\sfrac{1}{2}}^R)+F(U_{i-\sfrac{1}{2}}^L)\right) - \frac{c_{i-\sfrac{1}{2}}}{2}\left(U_{i-\sfrac{1}{2}}^R - U_{i-\sfrac{1}{2}}^L\right).
    \end{align*} 
    
    \item \textbf{Compute the source term, $S_i$} 
    \[S_i = \begin{bmatrix}
        0 \\ -K_R \frac{Q_i}{A_i}
    \end{bmatrix}.\] 

    \item \textbf{Assemble the spatial operator $\Phi$} \\
    The semi-discrete formulation for cell $i$ is given by
    \[\Phi(U_{i-2},... U_{i+2}) = -\frac{F|_{z+\sfrac{1}{2}}-F|_{z-\sfrac{1}{2}}}{\Delta z} + S_i.\]

    \item \textbf{Advance in time using the Adams-Bashforth scheme}
    \[U^{n+1} = U^n + \Delta t \left(\frac{3}{2}\Phi\left(U^n\right) - \frac{1}{2} \Phi \left(U^{n-1}\right) \right)\]
\end{enumerate}

\section{Boundary conditions} \label{sec: Boundary conditions}
Assuming the source term is small at the boundaries, the governing equations for the characteristic variables can be written as
\begin{align}
    \frac{\partial W_1}{\partial t}+\lambda_1 \frac{\partial W_1}{\partial z} = 0,  \qquad 
    \frac{\partial W_2}{\partial t}+\lambda_2 \frac{\partial W_2}{\partial z} = 0.
\end{align}
To capture the wave propagation correctly, the characteristic variables $W_1$ and $W_2$ are used to prescribe the boundary conditions. 

Linear extrapolation is commonly used to compute the characteristic variables at the next time step. Based on the method of characteristics, a first order approximation for the outgoing characteristic variables at the next time step $(n+1)$ at the boundaries can be written as
\begin{align*}
    W_2^{n+1}|_{z=0} = W_2^n|_{z=0-\lambda_2 \Delta t} \\
    W_1^{n+1}|_{z=L} = W_1^n|_{z=L-\lambda_1\Delta t}.
\end{align*}
Detailed applications of the above expressions at each boundary are clearly described in the following texts. In other words, the value of the outgoing characteristic variable at the boundary is obtained by tracing it back along its characteristic curve by a distance $\lambda\, \Delta t$ to find its value at the previous time step.

\subsection{Inlet Boundary Condition}
In practice, however, the inlet boundary conditions are usually specified in terms of physical quantities such as volumetric flow rate $Q_{in}(t)$ or pressure $P_{in}(t)$ instead of directly prescribing the characteristic variables $W_1$ and $W_2$. Incorporating these physical inlet boundary conditions in terms of $W_1$ and $W_2$ is achieved using equation \eqref{eqn:Riemann invariants - in A and Q}.

Suppose the inlet condition is given as a prescribed $Q_{in}$. In this case, the outgoing characteristic variable at the inlet $(z=0)$ for $W_2$ is updated using the first-order linear extrapolation and $W_1$ with equation \eqref{eqn:Riemann invariants - in A and Q}:
\begin{equation*}
\begin{aligned}
W_2^{n+1}|_{z=0} &= W_2^n|_{z=0} + \Delta t\left(\frac{W_2^n|_{z=0+\Delta z}-W_2^n|_{z=0}}{\Delta z}\right)(-\lambda_2|_{z=0})\\
W_1^{n+1}|_{z=0} &= -W_2^{n+1}|_{z=0} + 2\frac{Q_{in}}{A^n}   
\end{aligned}
\end{equation*}
where $A^n$ is the area at the inlet from the current time step. Then $W_1^{n+1}$ and $W_2^{n+1}$ are converted back to the physical variables $(A, Q)$ using the following relations:
\begin{equation}
\label{eqn:convert to A Q}
\begin{aligned}
A^{n+1}|_{z=0} &= \left(\frac{W_1^{n+1}|_{z=0}-W_2^{n+1}|_{z=0}}{4}\right)^4\left(\frac{\rho}{\beta}\right)^2 \\ 
Q^{n+1}|_{z=0} &= A^{n+1}|_{z=0}\;\frac{W_1^{n+1}|_{z=0}+W_2^{n+1}|_{z=0}}{2}.    
\end{aligned}
\end{equation}

Alternatively, if the inlet boundary condition is given in terms of $P_{in}$, a similar procedure is followed:
\begin{equation*}
    \begin{aligned}
        W_2^{n+1}|_{z=0} &= W_2^n|_{z=0} + \frac{W_2^n|_{z=0+\Delta z}-W_2^n|_{z=0}}{\Delta z}(-\lambda_2|_{z=0} \; \Delta t)\\
        W_1^{n+1}|_{z=0} &= W_2^{n+1}|_{z=0} + 8\sqrt{\frac{1}{2\rho}(P_{in}-P_{ext}+\beta\sqrt{A_0})}.
    \end{aligned}
\end{equation*}
Then, $W_1^{n+1}$ and $W_2^{n+1}$ can be converted back to $(A, Q)$ using a procedure similar to that of equation~\eqref{eqn:convert to A Q}.

\begin{figure} [h!]
    \centering
    \includegraphics[width=\linewidth]{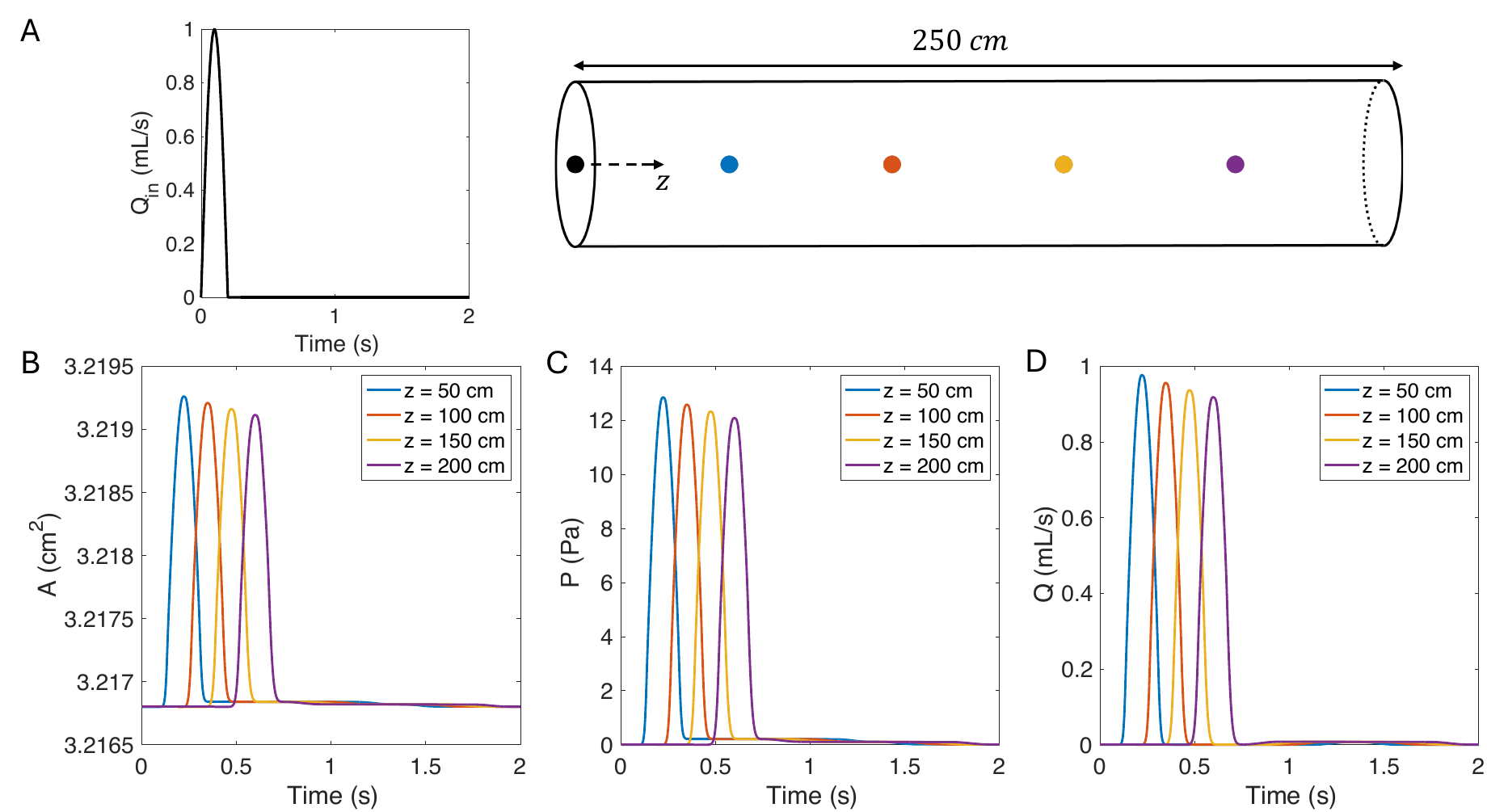}
    \caption{
    Simulation of a 1D single compliant vessel. (A) Representative inlet boundary condition in $Q$ that varies with time, along with an illustration of the compliant vessel. The black dot at the vessel indicates where the boundary condition is applied, and the color-coded nodes represent the positions $z=50, 100, 150$, and $250$ cm, respectively. (B) Propagation of the cross-sectional area $A$ in time and space. (C) Propagation of the internal pressure $P$ in time and space. (D) Propagation of the volumetric flow rate $Q$ in time and space.
    }
    \label{fig_result:inletQ results}
\end{figure}

The inlet boundary condition is typically imposed as a time-dependent waveform, specified either in $Q$ or $P$. Figure \ref{fig_result:inletQ results}A illustrates an example where the inlet boundary is defined as $Q_{in}(t)$. The solutions were numerically solved via the previously described MUSCL scheme. The results, which include the cross-sectional area $A$, the internal pressure $P$, and the volumetric flow rate $Q$ as functions of both space and time, are presented in Figure \ref{fig_result:inletQ results}B-D.

For this simulation, $Q_{in}$ is specified as 
\[Q_{in}=Q_c\cdot \sin(2\pi t/T_C)\cdot H(-t+T_C/2)\]
where $Q_C=1$ mL/s, $T_C=0.4$ s, and $H$ is the Heaviside function. The remaining parameters are set as follows: $L=250$ cm, $A_0=3.22$ cm$^2$, and $\beta=1.87*10^6$ Pa/m. These parameter values are adopted from \cite{wang20141d}.

\subsection{Outlet Boundary Condition}
Extrapolation of the characteristic variables is also used to handle the outlet boundary condition. Generally, there are two types of outlet boundary conditions used in 1D blood flow simulations: Non-reflective (or reflective) and Windkessel. The non-reflective boundary condition allows the outgoing characteristic to leave the computational domain without generating any spurious reflections. This approach is particularly useful when simulating vessel(s) without considering the effects of the peripheral circulation. On the other hand, the Windkessel boundary condition models the influence of the peripheral vasculature. Since it is computationally not feasible to simulate the entire arterial network, including small vessels and capillaries at downstream locations, the Windkessel model lumps the peripheral resistance and capacitance into a simplified representation. This approach allows analyzing the overall hemodynamic effects of the peripheral circulation at the outlet boundary.
\newline

\noindent {\normalsize \textbf{Non-reflecting/reflecting boundary conditions}}

Manipulating the reflections can be formulated in terms of the characteristic variables. Linear extrapolation is used to set these boundary conditions by extrapolating the outgoing characteristic variable and then adjusting the incoming variable with a reflection coefficient, $R_t$. For example, at the outlet $(z=L)$, the boundary conditions can be expressed as 
\begin{equation*}
    \begin{aligned}
        W_1^{n+1}|_{z=L} &= W_1^{n}|_{z=L} + \Delta t\left(\frac{W_1^n|_{z=L-\Delta z}-W_1^n|_{z=L}}{\Delta z}\right)\,\lambda_1|_{z=L}\\
        W_2^{n+1}|_{z=L} &= W_2^0|_{z=L} - R_t(W_1^{n+1}|_{z=L}-W_1^0|_{z=L})
    \end{aligned}
\end{equation*}
where $W_1^0$ and $W_2^0$ are the initial values of the Riemann invariants determined from the vessel properties, and $R_t$ is the reflection coefficient ranging from -1 to 1. When $R_t=0$, there is no reflection, meaning the outgoing wave leaves the computational domain without interference. Once the characteristic variables are updated at the boundary, the corresponding physical variables $(A,Q)$ are recovered by inverting the Riemann invariants using equation \eqref{eqn:convert to A Q}. 

\begin{figure} [h!]
    \centering
    \includegraphics[width=\linewidth]{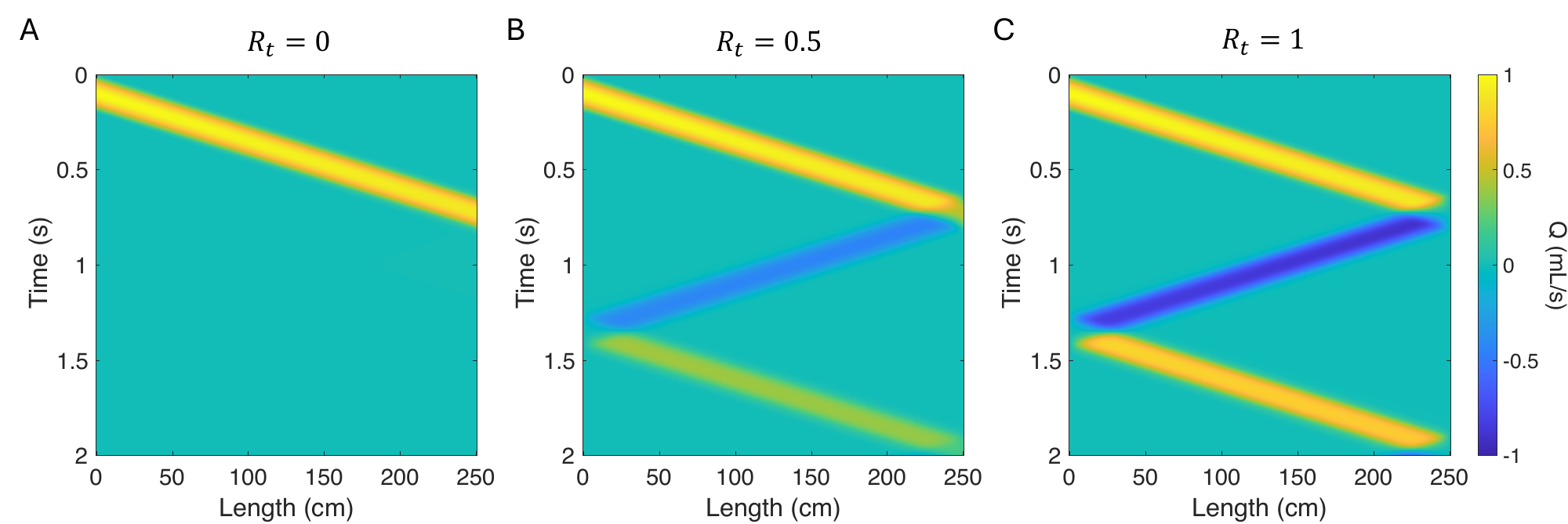}
    \caption{Representative example of the volume flow rate obtained using different reflection coefficients, $R_t$. (A) $R_t=0$ (no reflection), where the outgoing wave leaves the domain without interference; (B) $R_t=0.5$, indicating partial reflection; (C) $R_t=1$ (full reflection), where the reflected wave completely counterbalances the outgoing wave.}
    \label{fig_results: reflective}
\end{figure}

Figure \ref{fig_results: reflective} illustrates these non-reflective (Figure \ref{fig_results: reflective}A)/reflective (Figure \ref{fig_results: reflective}B-C) boundary conditions. As $R_t$ approaches 1, stronger reflections are observed at the boundaries, with the waveforms near the edges approaching zero due to the counterbalancing of two waves that are $180^{\circ}$ out of phase.
\newline 

\noindent {\normalsize \textbf{Windkessel boundary conditions}}

\begin{figure}
    \centering
    \includegraphics[width=0.8\linewidth]{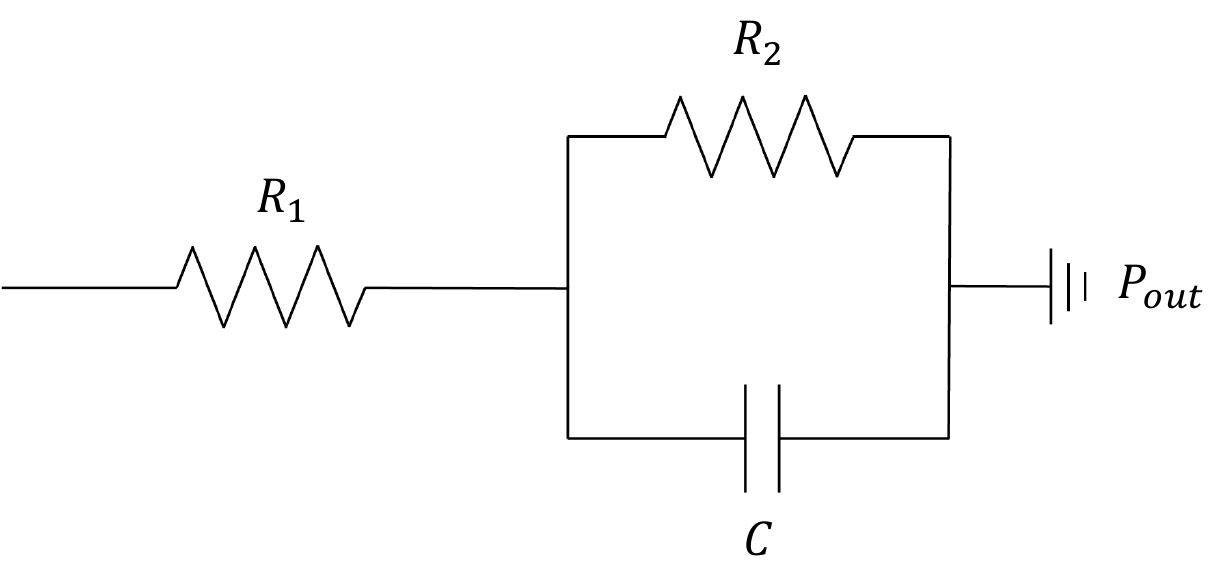}
    \caption{Representative figure of a 3-element Windkessel outlet boundary. $R_1$ is characteristic resistance determined by the properties of the proximal vessel, $R_2$ is the peripheral resistance, and $C$ is the peripheral compliance.}
    \label{fig:wk3}
\end{figure}

The explanation of the Windkessel boundary condition here mainly follows Alastruey et al.~\cite{alastruey2008lumped}. We first describe the model using the $(A,u)$ system as done in the original paper, and then convert the formulation to the $(A,Q)$ system. There are several types of Windkessel models, classified by the number of elements: 1-element ($R$), 2-elements ($LR$ or $RC$), 3-elements ($RCR$), and 4-elements: ($RCLR$). Here, $R$ represents resistance, $L$ represents inductance, and $C$ represents capacitance. Figure \ref{fig:wk3} illustrates the 3-element Windkessel outlet model.

The Windkessel concept is essentially a lumped parameter (0D) model of fluid flow that is analogous to an electrical circuit. In our explanation, we focus on the 3-elements Windkessel model, which consists of three elements: $R_1$, $R_2$, and $C$, described below.
\begin{adjustwidth}{1 cm}{1 cm}
\begin{itemize}
    \item \textbf{$R_1$}: This initial resistance is introduced to absorb approaching waves and reduce artificial reflection at the outlet. Its value is determined based on the properties of the proximal vessel. Specifically,
    \[
      R_1=\frac{\rho c_0}{A_0}
    \]
    where \( c_0=\sqrt{\frac{\beta}{2\rho}\sqrt{A_0}} \).
    \item \textbf{$R_2$}: This represents the resistance of the peripheral vasculature. Since \( R_1 \) is used to mitigate the wave reflections, \( R_2 \) is set as the remainder of the total resistance, i.e., 
    \[
      R_2 = \text{Total peripheral resistance}\; -\;R_1
    \]
    \item \textbf{$C$}: This is the capacitance of the periphery, representing its ability to store blood volume, similar to the way a capacitor stores charge.
\end{itemize}
\end{adjustwidth}
In addition, $P_{out}$ is the pressure at the end of the periphery (typically the venous pressure), which is usually assumed to be zero. The pressure across the capacitor is denoted by $P_C$. 

The coupling of the 1D blood flow domain to the lumped parameter 3-element Windkessel model is established by solving a Riemann problem at the interface between the 1D and 0D domains. In this approach, an intermediate state $(A^*, u^*)$ at the next time step is computed such that the outgoing characteristics from the 1D domain match the incoming characteristics of the Windkessel model. Specifically, if $(A_L, u_L)$ represents the state at the end of the 1D domain and $(A_R, u_R)$ represents the state at the start of the 0D domain, then the matching conditions are
\[W_1(A^*,u^*) = W_1(A_L, u_L) \qquad and \qquad W_2(A^*, u^*) = W_2(A_R, u_R).\]
Using the definitions of the Riemann invariants (as given in equation \eqref{eqn:Riemann invariants}), these conditions lead to a relation for the intermediate velocity:
\begin{equation}
\label{3WK: ustar}
    u^*=u_L + 4\sqrt{\frac{\beta}{2\rho}\sqrt{A_L}} - 4\sqrt{\frac{\beta}{2\rho}\sqrt{A^*}}.
\end{equation}
Furthermore, by inverting the Riemann invariants to recover the physical variables, one obtains
\begin{equation}
    \label{3WK: inverse W1 W2 to get Astar}
    A^* = \left[\frac{(W_1(A_L, u_L)-W_2(A_R, u_R))^4}{1024}\left(\frac{\rho}{\beta}\right)^2\right]
\end{equation}
\begin{equation}
    \label{3WK: inverse W1 W2 to get ustar}
    u^* = \frac{1}{2}(W_1(A_L,u_L)-W_2(A_R,u_R)).
\end{equation}
The coupled state $(A^*, u^*)$ then must satisfy the 3-element Windkessel model equations: 
\begin{equation}
\label{3WK: equation1}
    C\frac{dP_C}{dt}+\frac{P_C-P_{out}}{R_2}-A^*u^*=0
\end{equation}
\begin{equation}
\label{3WK: equation2}
    A^*u^* = \frac{P(A^*)-P_C}{R_1}
\end{equation}
with the constitutive relation
\[P(A^*) = P_{ext}+\beta(\sqrt{A^*}-\sqrt{A_0}).\]
\newline

\noindent \textbf{Steps for implementing 3-element Windkessel boundary condition}
\begin{enumerate}
    \item \textbf{Time discretization of the capacitor equation}\\
    Apply a first-order time discretization to equation \eqref{3WK: equation1} to update the capacitor pressure $P_C$ 
    \[P_C^n=P_C^{n-1}+\frac{\Delta t}{C}\left(A^* u^* - \frac{P_C^{n-1}-P_{out}}{R_2}\right)\]
    where the initial condition is assumed as $P_C^0=0$. 
    \item \textbf{Initial guess for $P_C^n$}\\
    Initially set $A^*=A_L^n$ and $u^*=u_L^n$ (values from the previous time step) to compute the first approximation of $P_C^n$ above. 
    \item \textbf{Formulate the nonlinear equation}\\
    Substitute the computed $P_C^n$ and the expression for $u^*$ from equation \eqref{3WK: ustar} into equation \eqref{3WK: equation2} to obtain a nonlinear equation:
    \[F = A^*\left(\left[u_L^n+4\sqrt{\frac{\beta}{2\rho}\sqrt{A_L^n}} \right] - 4\sqrt{\frac{\beta}{2\rho}\sqrt{A^*}} \right) - \left[\frac{(P_{ext}+\beta(\sqrt{A^*-A_0})-P_C^n}{R_1}\right]=0\] 
    \item \textbf{Solve for $A^*$}\\
    Solve the nonlinear equation $F=0$ for $A^*$. This can be achieved by using Newton's method with an initial guess of $A^*=A_L^n$, or by employing a nonlinear system solver such as \textit{fsolve} function in MATLAB (The Mathworks, inc., Version R2024a) or the function of the same name in Python's \textit{SciPy} library \cite{2020SciPy-NMeth}. 
    \item \textbf{Compute the inlet pressure of the 0D domain}\\
    Once $A^*$ is determined, compute the corresponding pressure at the 0D inlet using the constitutive equation.
    \item \textbf{Determine $u^*$}\\
    With $A^*$ and $P_C^n$ known, solve for $u^*$ from equation \eqref{3WK: equation2}:
    \[u^* = \frac{P(A^*)-P_C^n}{R_1A^*}\] 
    \item \textbf{Impose the 1D outflow boundary condition}\\
    The boundary condition at the 1D outflow can be imposed in one of two ways:
    \begin{itemize}
        \item Enforce $A_R^{n+1}=A_L^n$. When substituted into the inverse Riemann invariant equation \eqref{3WK: inverse W1 W2 to get ustar}, this yields
        \[u_R^{n+1}=2u^*-u_L^n\]
        \item Enforce $u_R^{n+1}=u_L^n$. When substituted into the inverse Riemann invariant equation \eqref{3WK: inverse W1 W2 to get Astar}, this yields
        \[A_R^{n+1}=\left[2(A^*)^{\sfrac{1}{4}}-(A_L^n)^{\sfrac{1}{4}}\right]^4\]
    \end{itemize}
    In our implementation, we chose to set first option. 
    \item \textbf{Compute the outlet flow rate}\\
    Finally, determine the outlet volumetric flow rate by:
    \[Q_{outlet}=A_{outlet}\; u_{outlet}\]
\end{enumerate}

\section{Conjunctions}\label{sec: Conjunctions}
Vessel branching is a common phenomenon in the arterial system. For instance, blood from the ascending aorta is distributed to the coronary artery, brachiocephalic artery, common carotid arteries, subclavian arteries, and the descending aorta. To accurately model such separations in 1D blood flow simulations, one must account for branching points, such as bifurcations (one parent vessel splitting into two daughter vessels), trifurcations (one parent vessel splitting into three three daughter vessels), or anastomoses (two parent vessels merge into one daughter vessel). 

In these cases, branching points typically coincide with discontinuities in material properties (e.g., $\beta$) or initial vessel area ($A_0$). Moreover, these junctions can be used to model topological changes due to external factors, such as alterations in $\beta$ following stent insertion or changes in $A_0$ due to aneurysm formation. 

The coupling at the branching points is accomplished by enforcing conservation of mass and momentum, along with matching the outgoing and incoming characteristics using the Riemann invariants. In practice, this means that the Riemann invariants extrapolated from the parent vessel are matched with those entering the daughter vessel(s). Let ($A_p$, $Q_p$) are the cross-sectional area and volumetric flow rate at the last cell of the parent vessel, and ($A_d$, $Q_d$) are the corresponding variables at the first cell of the daughter vessel. \\

\begin{figure}
    \centering
    \includegraphics[width=0.8\linewidth]{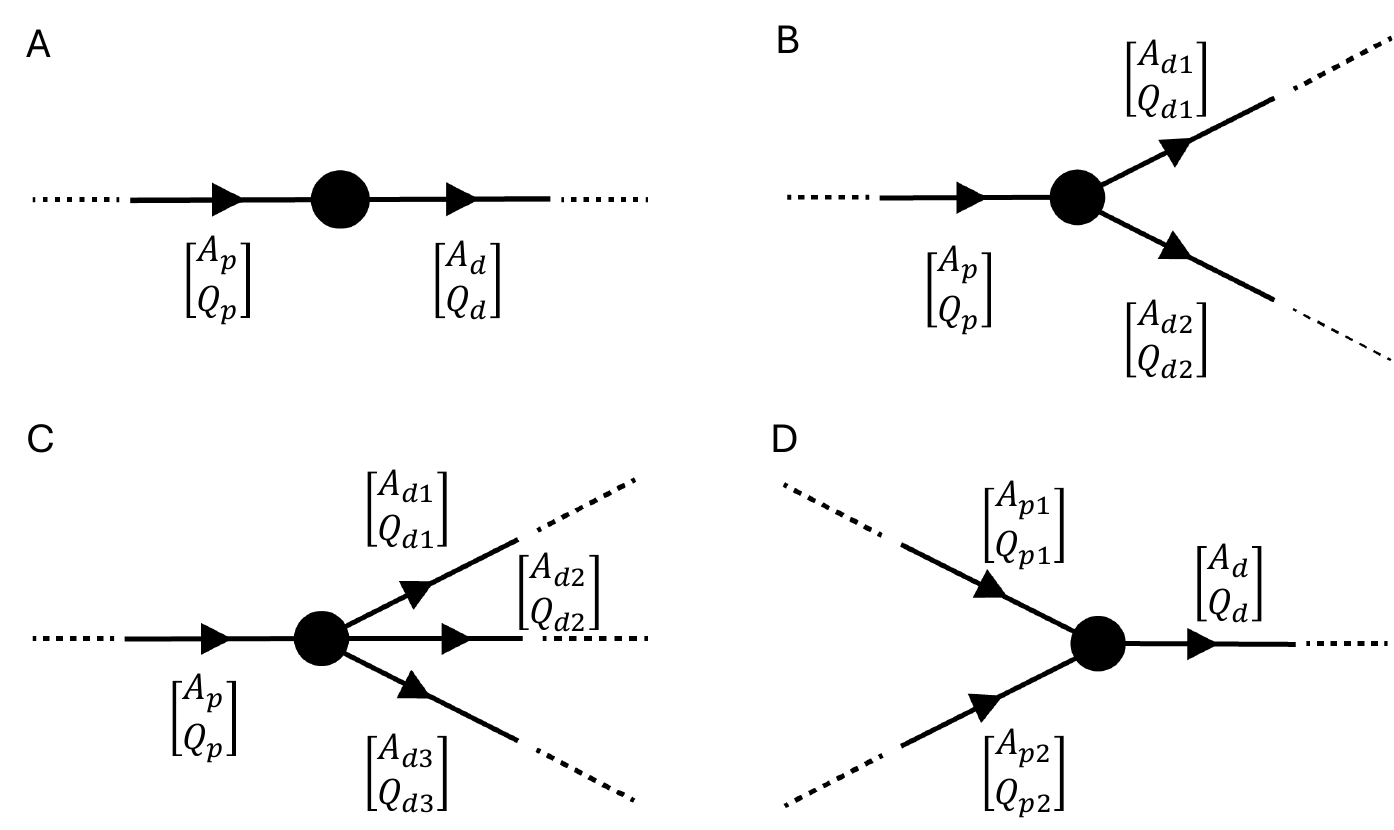}
    \caption{Representative examples of conjunction types: (A) a single conjunction, (B) a bifurcation, (C) a trifurcation, and (D) an anastomosis.}
    \label{fig:conjunction types}
\end{figure}

\noindent {\normalsize \textbf{Single conjunction}}

In the single conjunction case (illustrated in Figure \ref{fig:conjunction types}A), a parent vessel is connected to a daughter vessel that may have different properties. This example is presented in Formaggia et al.~\cite{formaggia2002one} and Sherwin et al.~\cite{sherwin2003one}, where a stent is placed in the middle of a vessel. The presence of the stent increases the local $\beta$ in the stented region, and this change in properties can be modeled by enforcing conservation laws and using characteristics matched at the junction.

The following equations describe a single conjunction coupling at the boundary of the two regions:
\begin{enumerate}
    \item \textbf{Conservation of Mass}\\
    The mass flux must be continuous across the junction:
    \[Q_p-Q_d=0\]
    \item \textbf{Conservation of Momentum}\\
    The total pressure (which includes the dynamic and static contributions) must be continuous:
    \begin{align*}
        \frac{1}{2}\rho \left(\frac{Q_p}{A_p}\right)^2+\left(P_{ext_p}+\beta_p\left(\sqrt{A_p}-\sqrt{A_{0_p}}\right)\right) = \frac{1}{2}\rho \left(\frac{Q_d}{A_d}\right)^2+\left(P_{ext_d}+\beta_d\left(\sqrt{A_d}-\sqrt{A_{0_d}}\right)\right)
    \end{align*}
    \item \textbf{Extrapolation of Riemann invariants (parent vessel)}\\
    Using the first-order approximation, 
    \[W_{1_p}^{n+1} = W_{1_p}^n|_{z=L_p}+\Delta t \frac{\left(W_{1_p}^n|_{z=L_p-\Delta z_p}-W_{1_p}^n|_{z=L_p}\right)}{\Delta z_p}\lambda_{1_p}^n|_{z=L_p} = \frac{Q_p}{A_p}+4\sqrt{\frac{\beta_p}{2\rho}\sqrt{A_p}}\]
    \item \textbf{Extrapolation of Riemann invariants (daughter vessels)}\\
    Similarly, the incoming characteristic variable is linearly extrapolated from the interior:
    \[W_{2_d}^{n+1} = W_{2_d}^n|_{z=0}+\Delta t \frac{\left(W_{2_d}^n|_{z=0+\Delta z_d}-W_{2_d}^n|_{z=0}\right)}{\Delta z_d}(-\lambda_{2_d}^n|_{z=0}) = \frac{Q_d}{A_d}-4\sqrt{\frac{\beta_d}{2\rho}\sqrt{A_d}}\]
\end{enumerate}
In these expressions, $W_1^n$ and $W_2^n$ represent the characteristic variables from the current time step, and the subscripts $p$ and $d$ refer to the parent and daughter vessels, respectively. Since there are four unknowns ($A_p$, $Q_p$, $A_d$, and $Q_d$) and four equations, this system can be solved using Newton's iterative method (by computing the Jacobian of the system) or by using a nonlinear system solver from MATLAB or Python. \\

\noindent {\normalsize \textbf{Bifurcation/trifurcation}} 

Bifurcation is one of the most common branching types in the arterial system. Its schematic in a 1D domain is illustrated in Figure \ref{fig:conjunction types}B. The modeling approach for a  bifurcation is similar to that for a single conjunction, but with the addition of a second daughter vessel. In this case, let the variables for the parent vessel at the junction be denoted by $(A_p, Q_p)$, while $(A_{d1}, Q_{d1})$ and $(A_{d2}, Q_{d2})$ refer to that of daughter vessel 1 and 2 respectively. The following set of equations governs the coupling at the bifurcation:
\begin{enumerate}
    \item \textbf{Conservation of mass}
    \[Q_p-Q_{d1}-Q_{d2}=0\]
    \item \textbf{Conservation of momentum (parent vessel $\rightarrow$ daughter vessel 1)}
    \begin{align*}
        \frac{1}{2}\rho \left(\frac{Q_p}{A_p}\right)^2+\left(P_{ext_p}+\beta_p\left(\sqrt{A_p}-\sqrt{A_{0_p}}\right)\right) = \frac{1}{2}\rho \left(\frac{Q_{d1}}{A_{d1}}\right)^2+\left(P_{ext_{d1}}+\beta_{d1}\left(\sqrt{A_{d1}}-\sqrt{A_{0_{d1}}}\right)\right)
    \end{align*}
    \item \textbf{Conservation of momentum (parent vessel $\rightarrow$ daughter vessel 2)}
    \begin{align*}
        \frac{1}{2}\rho \left(\frac{Q_p}{A_p}\right)^2+\left(P_{ext_p}+\beta_p\left(\sqrt{A_p}-\sqrt{A_{0_p}}\right)\right) = \frac{1}{2}\rho \left(\frac{Q_{d2}}{A_{d2}}\right)^2+\left(P_{ext_{d2}}+\beta_{d2}\left(\sqrt{A_{d2}}-\sqrt{A_{0_{d2}}}\right)\right)
    \end{align*}
    \item \textbf{Extrapolation of Riemann invariants (parent vessel)}
    \[W_{1_p}^{n+1} = W_{1_p}^n|_{z=L_p}+\Delta t \frac{\left(W_{1_p}^n|_{z=L_p-\Delta z_p}-W_{1_p}^n|_{z=L_p}\right)}{\Delta z_p}\lambda_{1_p}^n|_{z=L_p} = \frac{Q_p}{A_p}+4\sqrt{\frac{\beta_p}{2\rho}\sqrt{A_p}}\]
    \item \textbf{Extrapolation of Riemann invariants (daughter vessel 1)}
    \[W_{2_{d1}}^{n+1} = W_{2_{d1}}^n|_{z=0}+\Delta t \frac{\left(W_{2_{d1}}^n|_{z=0+\Delta z_{d1}}-W_{2_{d1}}^n|_{z=0}\right)}{\Delta z_{d1}}(-\lambda_{2_{d1}}^n|_{z=0}) = \frac{Q_{d1}}{A_{d1}}-4\sqrt{\frac{\beta_{d1}}{2\rho}\sqrt{A_{d1}}}\]
    \item \textbf{Extrapolation of Riemann invariants (daughter vessel 2)}
    \[W_{2_{d2}}^{n+1} = W_{2_{d2}}^n|_{z=0}+\Delta t \frac{\left(W_{2_{d2}}^n|_{z=0+\Delta z_{d2}}-W_{2_{d2}}^n|_{z=0}\right)}{\Delta z_{d2}}(-\lambda_{2_{d2}}^n|_{z=0}) = \frac{Q_{d2}}{A_{d2}}-4\sqrt{\frac{\beta_{d2}}{2\rho}\sqrt{A_{d2}}}\]
\end{enumerate}
Because there are six unknowns $(A_p, Q_p, A_{d1}, Q_{d1},A_{d2}, Q_{d2})$ and six equations, this system can be solved using the same method described previously. 

The modeling of trifurcation is analogous to the bifurcation case, with the addition of one more conservation of momentum between the parent vessel and the third daughter vessel, as well as an additional condition for Riemann invariants for the third branch. \\

\noindent {\normalsize \textbf{Anastomosis}}

Anastomosis is a case where two parent vessels merge into a single daughter vessel, and it is illustrated in Figure \ref{fig:conjunction types}D. This configuration is often observed in the Circle of Willis, which is a network of arteries at the base of the brain. The modeling of an anastomosis follows a similar approach to that described previously, but with the appropriate changes in the sign conventions to reflect the merging of flows. 

Let the state of end cells of the two parent vessels be denoted by $(A_{p1}, Q_{p1}$ and $(A_{p2}, Q_{p2})$, respectively. Let the state of the beginning cell of the daughter vessel be denoted by $(A_d, Q_d)$. The coupling at the junction is enforced by the following equations:
\begin{enumerate}
    \item \textbf{Conservation of mass}
    \[Q_{p1}+Q_{p2}-Q_{d}=0\]
    \item \textbf{Conservation of momentum (parent vessel 1 $\rightarrow$ daughter vessel)}
    \begin{align*}
        \frac{1}{2}\rho \left(\frac{Q_{p1}}{A_{p1}}\right)^2+\left(P_{ext_{p1}}+\beta_{p1}\left(\sqrt{A_{p1}}-\sqrt{A_{0_{p1}}}\right)\right) = \frac{1}{2}\rho \left(\frac{Q_{d}}{A_{d}}\right)^2+\left(P_{ext_{d}}+\beta_{d}\left(\sqrt{A_{d}}-\sqrt{A_{0_{d}}}\right)\right)
    \end{align*}
    \item \textbf{Conservation of momentum (parent vessel 2 $\rightarrow$ daughter vessel)}
    \begin{align*}
        \frac{1}{2}\rho \left(\frac{Q_{p2}}{A_{p2}}\right)^2+\left(P_{ext_{p2}}+\beta_{p2}\left(\sqrt{A_{p2}}-\sqrt{A_{0_{p2}}}\right)\right) = \frac{1}{2}\rho \left(\frac{Q_{d}}{A_{d}}\right)^2+\left(P_{ext_{d}}+\beta_{d}\left(\sqrt{A_{d}}-\sqrt{A_{0_{d}}}\right)\right)
    \end{align*}
    \item \textbf{Extrapolation of Riemann invariants (parent vessel 1)}
    \begin{align*}
        W_{1_{p1}}^{n+1} = W_{1_{p1}}^n|_{z=L_{p1}}+\Delta t \frac{\left(W_{1_{p1}}^n|_{z=L_{p1}-\Delta z_{p1}}-W_{1_{p1}}^n|_{z=L_{p1}}\right)}{\Delta z_{p1}}\lambda_{1_{p1}}^n|_{z=L} = \frac{Q_{p1}}{A_{p1}}+4\sqrt{\frac{\beta_{p1}}{2\rho}\sqrt{A_{p1}}}
    \end{align*}
    \item \textbf{Extrapolation of Riemann invariants (parent vessel 2)}
    \begin{align*}
        W_{1_{p2}}^{n+1} = W_{1_{p2}}^n|_{z=L_{p2}}+\Delta t \frac{\left(W_{1_{p2}}^n|_{z=L_{p2}-\Delta z_{p2}}-W_{1_{p2}}^n|_{z=L_{p2}}\right)}{\Delta z_{p2}}\lambda_{1_{p2}}^n|_{z=L_{p2}} = \frac{Q_{p2}}{A_{p2}}+4\sqrt{\frac{\beta_{p2}}{2\rho}\sqrt{A_{p2}}}
    \end{align*}
    \item \textbf{Extrapolation of Riemann invariants (daughter vessel)}
    \[W_{2_{d}}^{n+1} = W_{2_{d}}^n|_{z=0}+\Delta t \frac{\left(W_{2_{d}}^n|_{z=0+\Delta z_d}-W_{2_{d}}^n|_{z=0}\right)}{\Delta z_{d}}(-\lambda_{2_{d}}^n|_{z=0}) = \frac{Q_{d}}{A_{d}}-4\sqrt{\frac{\beta_{d}}{2\rho}\sqrt{A_{d}}}\]
\end{enumerate}
Since there are six unknowns and six equations, these unknowns can be solved iteratively. 

\section{Verification}
The verification of the presented explanations (numerical scheme, treatment of boundary conditions, and treatment of junctions) were performed by comparing our results with those reported by Boileau et al. \cite{boileau2015benchmark} (inspired by Melis's thesis \cite{melis2017gaussian}). In particular, our 1D blood flow simulation solver (based on the finite volume MUSCL scheme) was bench-marked against a 1D finite element method using a Locally Conservative Galerkin scheme (1D FEM), which is widely used in 1D blood flow modeling, as well as against a full 3D simulation from Boileau et al. Note that each plot was manually digitized from the original figures. 

\begin{figure}[t!]
    \centering
    \includegraphics[width=1\linewidth]{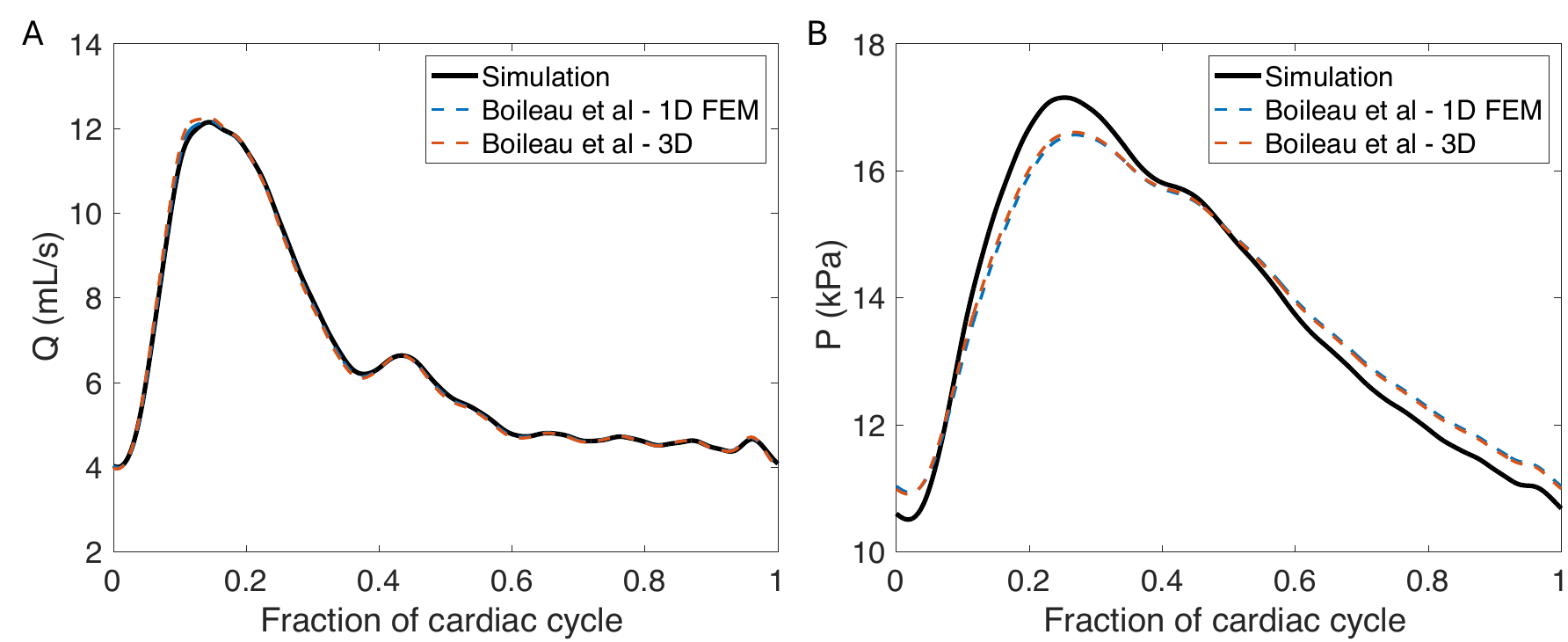}
    \caption{Verification of the presented simulation for a single common carotid artery. (A) Comparison of volume flow rate. (B) Comparison of internal pressures. In each plot, the black curve represents the simulation results from this study, which are compared to the benchmark data of Boileau et al.~\cite{boileau2015benchmark}: the blue dashed curve corresponds to the 1D FEM results and the orange dashed curve corresponds to the 3D simulation results, over one cardiac cycle.}
    \label{fig:CCA results}
\end{figure}

\begin{figure}
    \centering
    \includegraphics[width=0.9\linewidth]{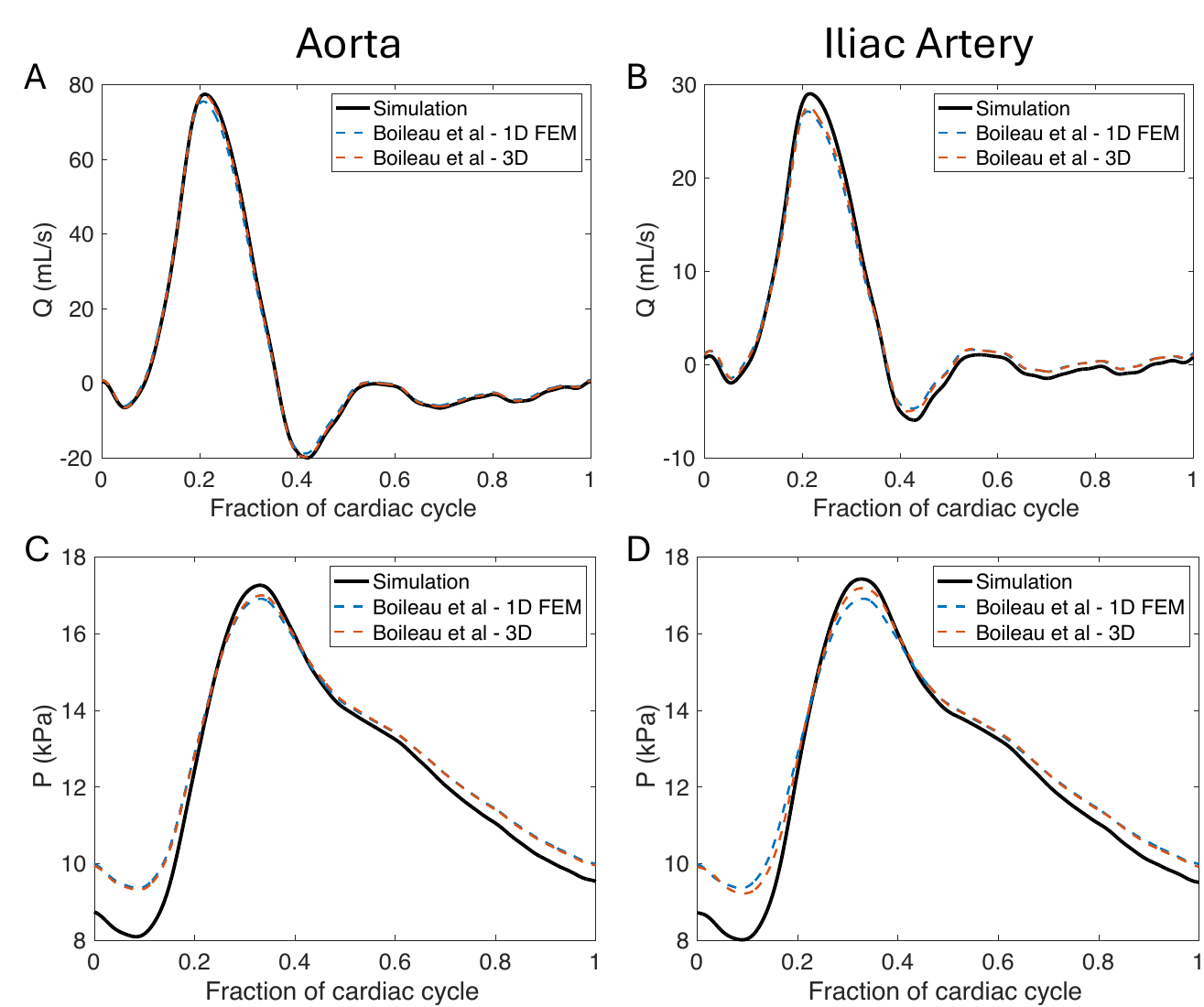}
    \caption{Verification of the presented simulation in the bifurcation case, where a single aorta bifurcates into two identical iliac arteries. (A) Volume flow rate in the aorta, (B) volume flow rate in the iliac artery, (C) internal pressure in the aorta, and (D) internal pressure in the iliac artery, each over one cardiac cycle.}
    \label{fig:bifurcation results}
\end{figure}

The tests were carried out for two cases: A single common carotid artery (Figure \ref{fig:CCA results}), and a bifurcation case where a single aorta bifurcates into two identical iliac arteries (Figure \ref{fig:bifurcation results}). In both cases, all outlets were coupled to a 3-element non-reflecting Windkessel boundary condition (characteristic $R_1$). The simulation parameters for these cases are summarized in Tables \ref{tab:CCA} and \ref{tab:bifurcation}.

Overall, the results from the presented 1D blood flow simulations match well with the benchmark data both qualitatively and quantitatively. The slight deviations observed in internal pressure in Figures \ref{fig:CCA results}B, \ref{fig:bifurcation results}C-D are likely due to differences in the constitutive equations that relate internal pressure to the cross-sectional area. The benchmark study used the relation $P=P_{ext}+P_d+\frac{\beta}{A_d}(\sqrt{A}-\sqrt{A_d})$, where $P_d$ and $A_d$ denote the diastolic pressure and area, respectively. Such differences influence the Moens-Korteweg celerity $c$ and the characteristic resistance $R_1$, thereby causing small discrepancies in the internal pressure. 

\begin{table}[h!]
    \centering
    \caption{List of parameters used in the single common carotid artery simulation taken from \cite{xiao2014systematic}.}
    \begin{tabular}{ccc}
        \hline
        \textbf{Property} & \textbf{Value} & \textbf{Units} \\
        \hline
        Length, $L$ & 126 & $\mathrm{mm}$ \\
        Initial cross-sectional area, $A_0$ & 0.220 & $\mathrm{cm}^2$ \\
        Vessel wall thickness, $h$ & 0.30 & $\mathrm{mm}$ \\
        Young's modulus, $E$ & 700 & $\mathrm{kPa}$ \\
        Blood density, $\rho$ & 1060 & $\mathrm{kg/m^3}$ \\
        Blood viscosity, $\mu$ & 4 & $\mathrm{mPa\cdot s}$ \\
        Velocity Profile correction factor, $\gamma$ & 2 & - \\
        External pressure, $P_{ext}$ & 0 & $\mathrm{Pa}$ \\
        Outlet pressure, $P_{out}$ & 0 & $\mathrm{Pa}$ \\
        Windkessel resistance, $R_1$ & $3.38 \times 10^8$ & $\mathrm{Pa \cdot s /m^3}$ \\
        Windkessel resistance, $R_2$ & $1.78\times 10^9$ & $\mathrm{Pa \cdot s /m^3}$ \\
        Windkessel capacitance, $C$ & $1.75 \times 10^{-10}$ & $\mathrm{m^3/Pa}$ \\
        \hline
    \end{tabular}
    \label{tab:CCA}
\end{table}

\begin{table}[h!]
    \centering
    \caption{List of parameters used for the bifurcation test in which the single aorta splits into two identical iliac arteries taken from \cite{xiao2014systematic}.}
    \begin{tabular}{ccc}
        \hline
        \textbf{Property} & \textbf{Aorta} & \textbf{Iliac} \\
        \hline
        Length, $L$ & 8.6 $\mathrm{cm}$ & 8.5 $\mathrm{cm}$ \\
        Initial cross-sectional area, $A_0$ & 0.86 $\mathrm{cm^2}$ & 0.60 $\mathrm{cm^2}$ \\
        Vessel wall thickness, $h$ & 1.03 $\mathrm{mm}$ & 0.72 $\mathrm{mm}$\\
        Young's modulus, $E$ & 500 $\mathrm{kPa}$ & 700 $\mathrm{kPa}$ \\
        Blood density, $\rho$ & 1060 $\mathrm{kg/m^3}$ & 1060 $\mathrm{kg/m^3}$ \\
        Blood viscosity, $\mu$ & 4 $\mathrm{mPa\cdot s}$ & 4 $\mathrm{mPa\cdot s}$ \\
        Velocity Profile correction factor, $\gamma$ & 2 & 2 \\
        External pressure, $P_{ext}$ & 0 $\mathrm{Pa}$ & 0 $\mathrm{Pa}$ \\
        Outlet pressure, $P_{out}$ & 0 $\mathrm{Pa}$ & 0 $\mathrm{Pa}$ \\
        Windkessel resistance, $R_1$ & - & $8.46 \times 10^7$ $\mathrm{Pa \cdot s/m^3}$ \\
        Windkessel resistance, $R_2$ & - & $3.08 \times 10^9$ $\mathrm{ Pa \cdot s/m^3}$\\
        Windkessel capacitance, $C$ & - & $3.67\times 10^{-10}$  $ \mathrm{m^3/Pa}$ \\
        \hline
    \end{tabular}
    \label{tab:bifurcation}
\end{table}

\section{Applications}
\begin{figure}[H]
    \centering
    \includegraphics[width=0.9\linewidth]{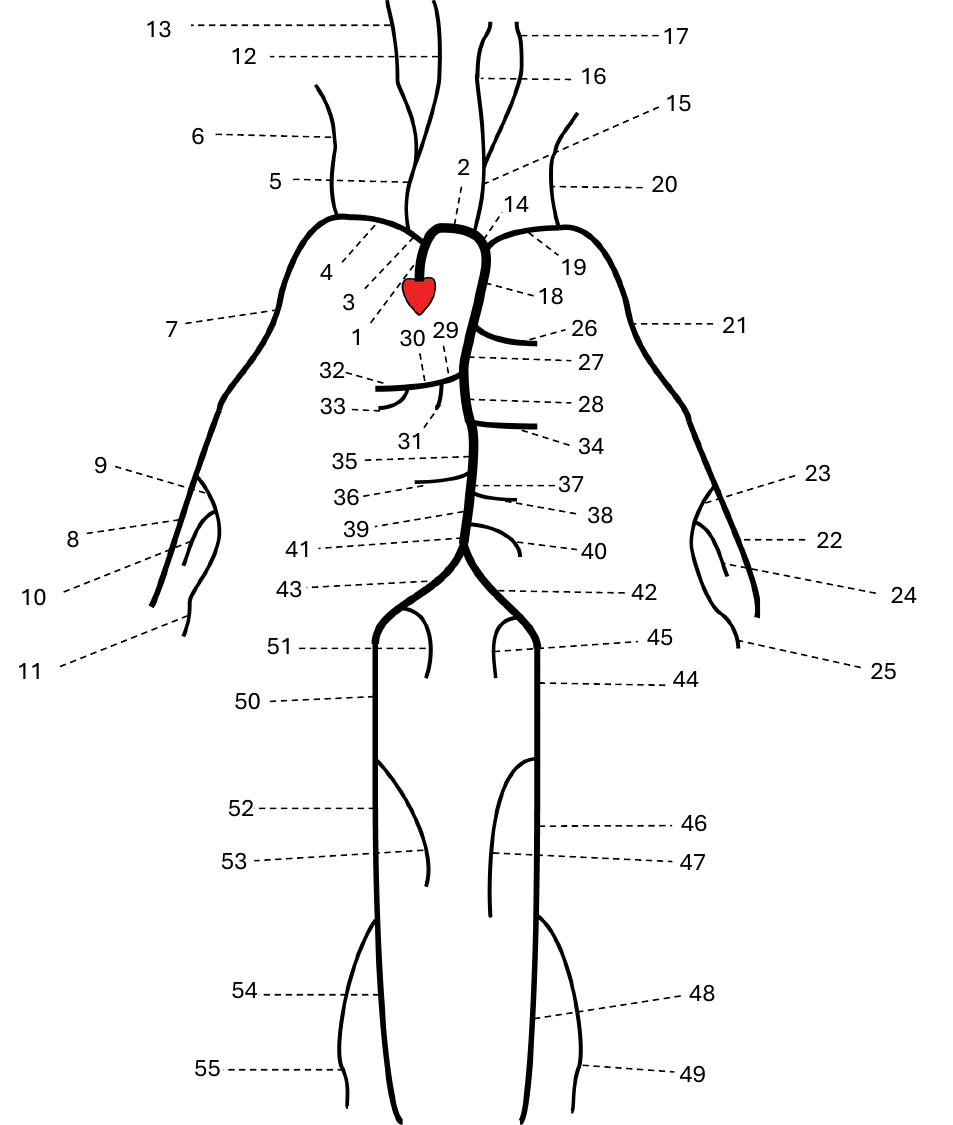}
    \caption{Illustration of the major arteries in the human body. Each numbered segment corresponds to an entry in Table~\ref{tab:full artery}. This figure was recreated based on \cite{wang2015verification}.}
    \label{fig:full artery}
\end{figure}

\begin{figure}[H]
    \centering
    \includegraphics[width=1\linewidth]{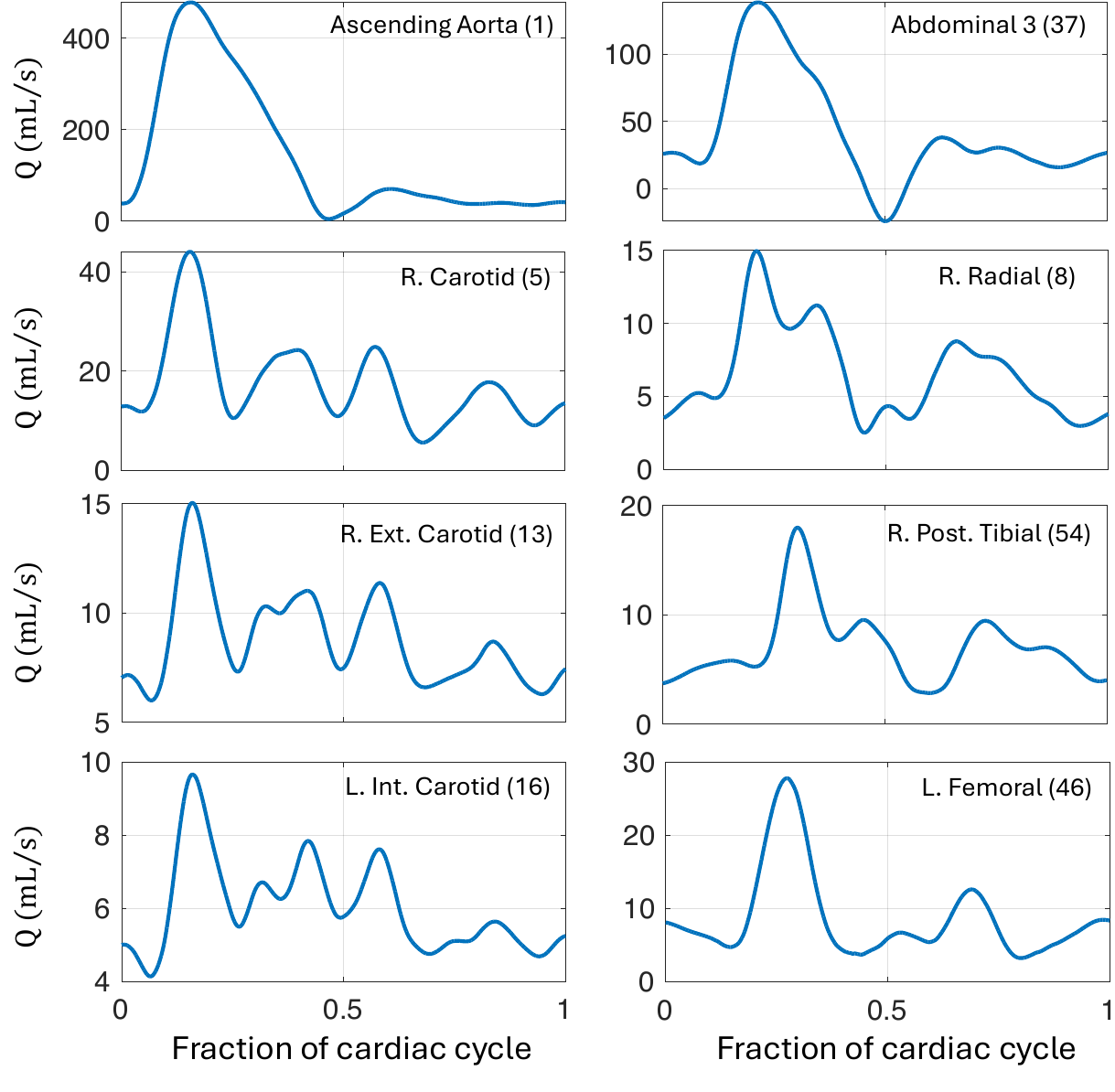}
    \caption{Results of representative volume flow rates in the full systemic arterial network.}
    \label{fig:full_artery_result}
\end{figure}

\begin{figure}[H]
    \centering
    \includegraphics[width=0.7\linewidth]{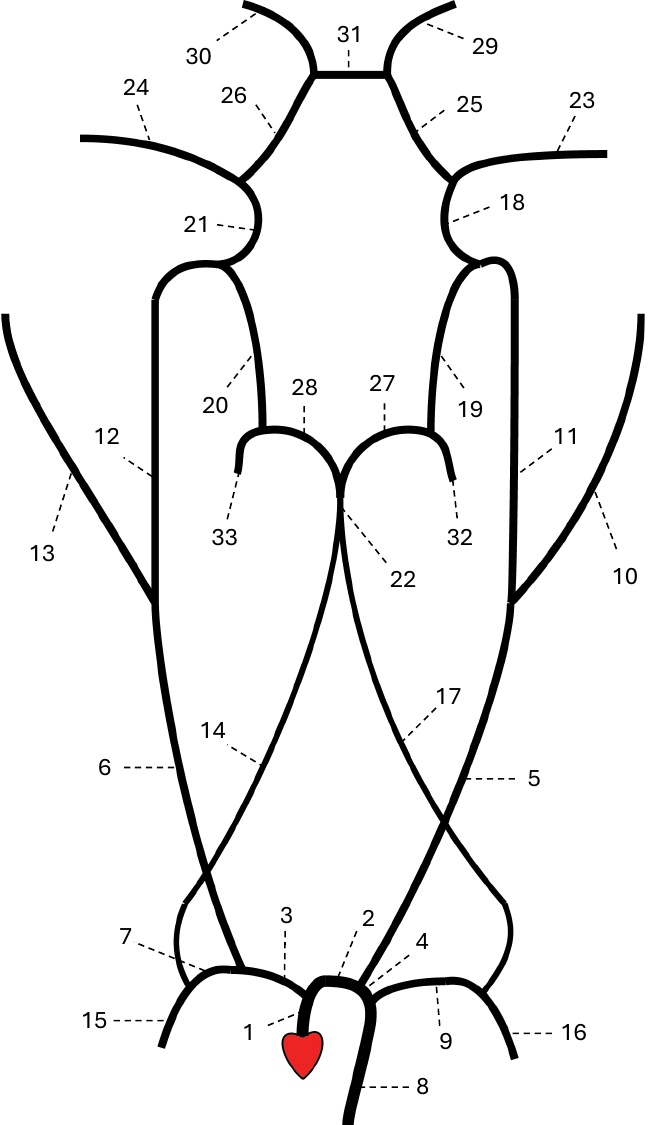}
    \caption{Illustration of major arteries that constitute the circle of Willis. Each numbered segment corresponds to an entry in Table~\ref{tab:circle of willis}. This figure was recreated based on \cite{alastruey2007modelling}.}
    \label{fig:circle of willis}
\end{figure}

\begin{figure}[H]
    \centering
    \includegraphics[width=1\linewidth]{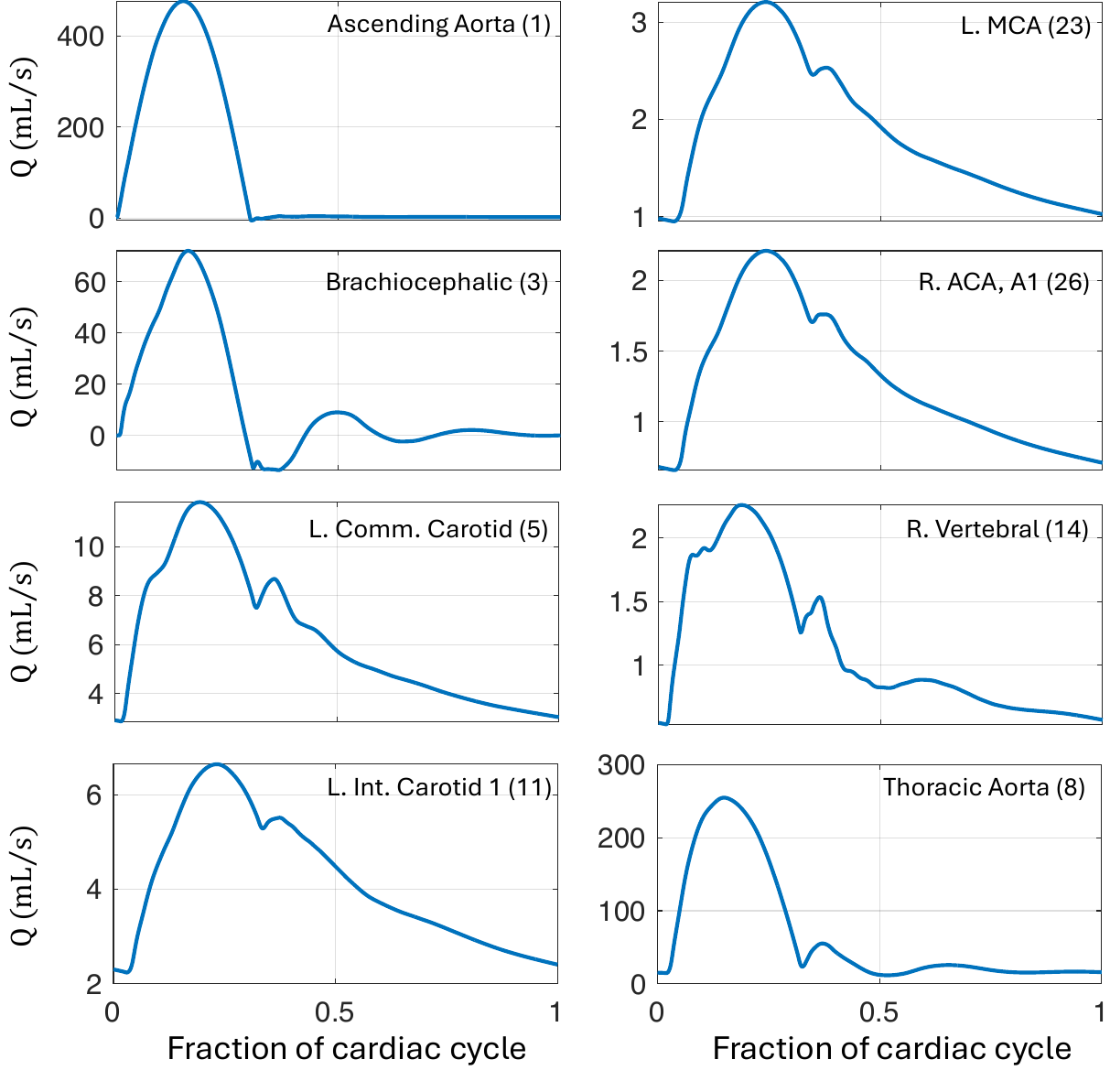}
    \caption{Results of representative volume flow rates in the circle of Willis arteries.}
    \label{fig:circle_of_willis_results}
\end{figure}

One of the main advantages of the reduced-order modeling is its computational efficiency, which makes it well-suited for modeling extensive arterial networks. In this section, we present two examples: a full systemic arterial network and the circle of Willis. The schematic of the full systemic arterial network is shown in Figure \ref{fig:full artery}, and the corresponding parameters are summarized in Table \ref{tab:full artery}. Representative volumetric flow rate profiles for this network are provided in Figure \ref{fig:full_artery_result}, with each plot labeled accordingly. In this simulation, the outlet boundary conditions were treated using reflective boundary conditions~\cite{sherwin2003computational}. Similarly, the schematic of the circle of Willis is presented in Figure~\ref{fig:circle of willis}, with its parameters summarized in Table~\ref{tab:circle of willis}. Representative volumetric flow rate profiles of the circle of Willis are shown in Figure~\ref{fig:circle_of_willis_results}, with each plot labeled accordingly. In this case, the outlet boundary conditions were implemented using 3-element Windkessel models. The steps for implementing either network system are as follows:
\begin{enumerate}
    \item \textbf{Set up inlet boundary conditions}\\
    Set up inlet boundary conditions in either $P(t)$ or $Q(t)$. 
    \item \textbf{Define vessel properties}\\
    For each vessel segment, specify the segment length, initial area ($A_0$), material property parameter $(\beta$), and the number of computational cells. 
    \item \textbf{Setup the outlet boundary condition}\\
    For each terminal segment, assign the outlet boundary conditions using either a reflection coefficient ($R_t$) or 3-element Windkessel properties (i.e., $R_1$, $R_2$, and $C$). 
    \item \textbf{Configure junctions (conjunctions)}\\
    Specify the configurations at the vessel junctions, ensuring that the connecting relationship between mother and daughter vessels are defined. 
    \item \textbf{Start the simulation time loop} 
    \item \textbf{Apply the inlet boundary condition}\\
    Apply the inlet boundary condition on the first cell of the first segment (e.g., the ascending aorta). 
    \item \textbf{Solve for the variables $(A, Q)$ in the current vessel segment}\\
    Use the numerical scheme to solve for the variables within a given vessel segment. 
    \item \textbf{Apply junction conditions}\\
    At each junction, apply and solve the conjunction equations that relate the end cell of the mother vessel(s) to the first cell of the daughter vessel(s). 
    \item \textbf{Iterate through all segments}\\
    Repeat steps 7 and 8 until the variables in every segment of the network are computed. 
    \item \textbf{Apply outlet boundary conditions}\\
    For each terminal segment, enforce the designated outlet boundary conditions. 
    \item \textbf{Advance in time}\\
    Return to step 5 and continue the time-stepping loop until the total simulation time reaches the specified \textit{end} time
\end{enumerate}

\subsection{Full Systemic Arterial Network}

\begin{longtable}{cccccc}
  \caption{List of parameters used in the full systemic arterial network taken from \cite{sherwin2003computational}.}
  \label{tab:full artery} \\
  \hline
  Number & Name & Length $\mathrm{(cm)}$ & $A_0\ \mathrm{(cm^2)}$ & $\beta\ (10^8\,\mathrm{Pa/m})$ & $R_t$ \\
  \hline
  \endfirsthead
  \multicolumn{6}{c}{\tablename\ \thetable\ -- \textit{Continued from previous page}} \\
  \hline
  Number & Name & Length $\mathrm{(cm)}$ & $A_0\ \mathrm{(cm^2)}$ & $\beta\ (10^8\,\mathrm{Pa/m})$ & $R_t$ \\
  \hline
  \endhead
  \hline \multicolumn{6}{r}{\textit{Continued on next page}} \\
  \endfoot
  \hline
  \endlastfoot
        1 & Ascending Aorta & 4.0 & 6.789 & 0.023 & - \\
        2 & Aortic Arch \MakeUppercase{\romannumeral 1} & 2.0 & 5.011 & 0.024 & - \\
        3 & Brachiocephalic & 3.4 & 1.535 & 0.049 & - \\
        4 & R. Subclavian \MakeUppercase{\romannumeral 1} & 3.4 & 0.919 & 0.069 & - \\
        5 & R. Carotid & 17.7 & 0.703 & 0.085 & - \\
        6 & R. Vertebral & 14.8 & 0.181 & 0.047 & 0.906 \\
        7 & R. Subclavian \MakeUppercase{\romannumeral 2} & 42.2 & 0.833 & 0.076 & - \\
        8 & R. Radius & 23.5 & 0.423 & 0.192 & 0.820 \\
        9 & R. Ulnar \MakeUppercase{\romannumeral 1} & 6.7 & 0.648 & 0.134 & - \\
        10 & R. Interosseous & 7.9 & 0.118 & 0.895 & 0.956 \\
        11 & R. Ulnar \MakeUppercase{\romannumeral 2} & 17.1 & 0.589 & 0.148 & 0.893 \\
        12 & R. Int. Carotid & 17.6 & 0.458 & 0.186 & 0.784 \\
        13 & R. Ext. Carotid & 17.7 & 0.458 & 0.173 & 0.790 \\
        14 & Aortic arch \MakeUppercase{\romannumeral 2} & 3.9 & 4.486 & 0.024 & - \\
        15 & L. Carotid & 20.8 & 0.536 & 0.111 & - \\
        16 & L. Int Carotid & 17.6 & 0.350 & 0.243 & 0.784 \\
        17 & L. Ext. Carotid & 17.7 & 0.350 & 0.227 & 0.791 \\
        18 & Thoracic Aorta \MakeUppercase{\romannumeral 1} & 5.2 & 3.941 & 0.026 & - \\
        19 & L. Subclavian \MakeUppercase{\romannumeral 1} & 3.4 & 0.706 & 0.088 & - \\
        20 & L. Vertebral & 14.8 & 0.129 & 0.657 & 0.906 \\
        21 & L. Subclavian \MakeUppercase{\romannumeral 2} & 42.2 & 0.650 & 0.097 & - \\
        22 & L. Radius & 23.5 & 0.330 & 0.247 & 0.821 \\
        23 & L. Ulnar \MakeUppercase{\romannumeral 1} & 6.7 & 0.505 & 0.172 & - \\
        24 & L. Interosseous & 7.9 & 0.093 & 1.139 & 0.956 \\
        25 & L. Ulnar \MakeUppercase{\romannumeral 2} & 17.1 & 0.461 & 0.189 & 0.893 \\
        26 & Intercostals & 8.0 & 0.316 & 0.147 & 0.627 \\
        27 & Thoracic Aorta \MakeUppercase{\romannumeral 2} & 10.4 & 3.604 & 0.026 & - \\
        28 & Abdominal Aorta \MakeUppercase{\romannumeral 1} & 5.3 & 2.659 & 0.032 & - \\
        29 & Celiac \MakeUppercase{\romannumeral 1} & 2.0 & 1.086 & 0.056 & - \\
        30 & Celiac \MakeUppercase{\romannumeral 2} & 1.0 & 0.126 & 0.481 & - \\
        31 & Hepatic & 6.6 & 0.659 & 0.070 & 0.925 \\
        32 & Gastric & 7.1 & 0.442 & 0.096 & 0.921 \\
        33 & Splenic & 6.3 & 0.468 & 0.109 & 0.93 \\
        34 & Sup. Mesenteric & 5.9 & 0.782 & 0.083 & 0.934 \\
        35 & Abdominal Aorta \MakeUppercase{\romannumeral 2} & 1.0 & 2.233 & 0.034 & - \\
        36 & L. Renal & 3.2 & 0.385 & 0.130 & 0.861 \\
        37 & Abdominal Aorta \MakeUppercase{\romannumeral 3} & 1.0 & 1.981 & 0.038 & - \\
        38 & R. Renal & 3.2 & 0.385 & 0.130 & 0.861 \\
        39 & Abdominal Aorta \MakeUppercase{\romannumeral 4} & 10.6 & 1.389 & 0.051 & - \\
        40 & Inf. Mesenteric & 5.0 & 0.118 & 0.344 & 0.918 \\
        41 & Abdominal Aorta \MakeUppercase{\romannumeral 5} & 1.0 & 1.251 & 0.049 & - \\
        42 & R. Com. Iliac & 5.9 & 0.694 & 0.082 & - \\
        43 & L. Com. Iliac & 5.8 & 0.694 & 0.082 & - \\
        44 & L. Ext. Iliac & 14.4 & 0.730 & 0.137 & - \\
        45 & L. Int. Iliac & 5.0 & 0.285 & 0.531 & 0.925 \\
        46 & L. Femoral & 44.3 & 0.409 & 0.231 & - \\
        47 & L. Deep Femoral & 12.6 & 0.398 & 0.223 & 0.885 \\
        48 & L. Post. Tibial & 32.1 & 0.444 & 0.383 & 0.724 \\
        49 & L. Ant. Tibial & 34.3 & 0.123 & 1.197 & 0.716 \\
        50 & R. Ext. Iliac & 14.5 & 0.730 & 0.137 & - \\
        51 & R. Int. Iliac & 5.0 & 0.285 & 0.531 & 0.925 \\
        52 & R. Femoral & 44.4 & 0.409 & 0.231 & - \\
        53 & R. Deep Femoral & 12.7 & 0.398 & 0.223 & 0.888 \\
        54 & R. Post. Tibial & 32.2 & 0.442 & 0.385 & 0.724 \\
        55 & R. Ant. Tibial & 34.4 & 0.122 & 1.210 & 0.716 \\
\end{longtable}

\subsection{Circle of Willis}

\small \begin{longtable}{cccccccc}
  \caption{List of parameters used in the circle of Willis taken from \cite{alastruey2007modelling}.}
  \label{tab:circle of willis} \\
  \hline
  Number & Name & \makecell{Length\\(cm)} & \makecell{Initial\\radius\\(cm)}
       & \makecell{Wall\\thickness\\(cm)} & \makecell{Elastic\\modulus\\($10^6$\,Pa)}
       & \makecell{Peripheral\\Resistance\\(10$^9$ Pa$\cdot$s/m$^3$)} & \makecell{Peripheral\\Compliance\\(10$^{-10}$ m$^3$/Pa)} \\
  \hline
  \endfirsthead
  \multicolumn{8}{c}{\tablename\ \thetable\ -- \textit{Continued from previous page}} \\
  \hline
  Number & Name & \makecell{Length\\(cm)} & \makecell{Initial\\radius\\(cm)}
       & \makecell{Wall\\thickness\\(cm)} & \makecell{Elastic\\modulus\\(10$^6$ Pa)}
       & \makecell{Peripheral\\Resistance\\(10$^9$ Pa$\cdot$s/m$^3$)} & \makecell{Peripheral\\Compliance\\(10$^{-10}$ m$^3$/Pa)} \\
  \hline
  \endhead
  \hline \multicolumn{8}{r}{\textit{Continued on next page}} \\
  \endfoot
  \hline
  \endlastfoot
    1 & Ascending Aorta & 4.0 & 1.2 & 0.163 & 0.4 & - & - \\
    2 & Aortic Arch \MakeUppercase{\romannumeral 1} & 2.0 & 1.120 & 0.126 & 0.4 & - & - \\
    3 & Brachiocephalic & 3.4 & 0.620 & 0.080 & 0.4 & - & - \\
    4 & Aortic Arch \MakeUppercase{\romannumeral 2} & 3.9 & 1.070 & 0.115 & 0.4 & - & - \\
    5 & L. Common Carotid & 20.8 & 0.250 & 0.063 & 0.4 & - & - \\
    6 & R. Common Carotid & 17.7 & 0.250 & 0.063 & 0.4 & - & - \\
    7 & R. Subclavian & 3.4 & 0.423 & 0.067 & 0.4 & - & - \\
    8 & Thoracic Aorta & 15.6 & 0.999 & 0.110 & 0.4 & 0.18 & 38.70 \\
    9 & L. Subclavian & 3.4 & 0.423 & 0.067 & 0.4 & - & - \\
    10 & L. Ext. Carotid & 17.7 & 0.150 & 0.038 & 0.8 & 5.43 & 1.27 \\
    11 & L. Int. Carotid & 17.7 & 0.200 & 0.050 & 0.8 & - & - \\
    12 & R. Int. Carotid & 17.7 & 0.200 & 0.050 & 0.8 & - & - \\
    13 & R. Ext. Carotid & 17.7 & 0.150 & 0.038 & 0.8 & 5.43 & 1.27 \\
    14 & R. Vertebral & 14.8 & 0.136 & 0.034 & 0.8 & - & - \\
    15 & R. Brachial & 42.2 & 0.403 & 0.067 & 0.4 & 2.68 & 2.58 \\
    16 & L. Brachial & 42.2 & 0.403 & 0.067 & 0.4 & 2.68 & 2.58 \\
    17 & L. Vertebral & 14.8 & 0.136 & 0.034 & 0.8 & - & - \\
    18 & L. Int. Carotid \MakeUppercase{\romannumeral 2} & 0.5 & 0.200 & 0.050 & 1.6 & - & - \\
    19 & L. PCoA & 1.5 & 0.073 & 0.018 & 1.6 & - & - \\
    20 & R. PCoA & 1.5 & 0.073 & 0.018 & 1.6 & - & - \\
    21 & R. Int. Carotid \MakeUppercase{\romannumeral 2} & 0.5 & 0.200 & 0.050 & 1.6 & - & - \\
    22 & Basilar & 2.9 & 0.162 & 0.040 & 1.6 & - & - \\
    23 & L. MCA & 11.9 & 0.143 & 0.036 & 1.6 & 5.97 & 1.16 \\
    24 & R. MCA & 11.9 & 0.143 & 0.036 & 1.6 & 5.97 & 1.16 \\
    25 & L. ACA, A1 & 1.2 & 0.117 & 0.029 & 1.6 & - & - \\
    26 & R. ACA, A1 & 1.2 & 0.117 & 0.029 & 1.6 & - & - \\
    27 & L. PCA, P1 & 0.5 & 0.107 & 0.027 & 1.6 & - & - \\
    28 & R. PCA, P1 & 0.5 & 0.107 & 0.027 & 1.6 & - & - \\
    29 & L. ACA, A2 & 10.3 & 0.120 & 0.030 & 1.6 & 8.48 & 0.82 \\
    30 & R. ACA, A2 & 10.3 & 0.120 & 0.030 & 1.6 & 8.48 & 0.82 \\
    31 & ACoA & 0.3 & 0.074 & 0.019 & 1.6 & - & - \\
    32 & L. PCA, P2 & 8.6 & 0.105 & 0.026 & 1.6 & 11.08 & 0.62 \\
    33 & R. PCA, P2 & 8.6 & 0.105 & 0.026 & 1.6 & 11.08 & 0.62 \\
\end{longtable}

\section{Conclusion}
In this work, we present a comprehensive overview of 1D blood flow simulation. Our discussion includes a derivation of the governing equations, the formulation of a common numerical scheme, and an in-depth explanation of the method of characteristics, with a particular focus on the treatment of boundary conditions and junctions. Although numerous studies have utilized 1D blood flow models, few provide a step-by-step detailed guide. Our goal is to empower researchers with the tools and understanding needed to build their own robust simulations.

It is important to recognize that 1D blood flow models have inherent limitations. Unlike 2D or 3D simulations, 1D models provide only transversely averaged information along the vessel axis, and thus only capture detailed spatial variations in the axial direction. As a result, these models cannot resolve fine-scale features such as local wall shear stress variations \cite{ladisa2006alterations}, oscillatory shear index \cite{numata2016blood}, vorticity \cite{schenkel2009mri}, helicity \cite{morbiducci2007helical}, or intricate energy loss patterns \cite{pekkan2005physics}, which are biomarkers commonly used in computational fluid dynamics applications in the cardiovascular system. Additionally, 1D simulations do not inherently capture the effect of vessel curvature or the complex rheological properties of blood without special treatment to capture these effects~\cite{formaggia2010cardiovascular, ghigo2018time}. Despite these limitations, the simplicity and efficiency of 1D models make them invaluable for applications where large-scale simulations are required. Furthermore, it is important to note that there are many alternative approaches for solving hyperbolic equations, setting boundary conditions \cite{vignon2010outflow}, and improving computational efficiency, such as through the integration of neural network \cite{csala2025physics}. The methodology presented here represents only one viable strategy among many.

We hope that this work serves as a valuable resource and foundation for both novice and experienced researchers. By providing clear, detailed guidance on the implementation of 1D blood flow simulations, our aim is to foster the development of efficient and accurate modeling tools that can be applied to a wide range of studies with applications in engineering, fundamental science, and clinical medicine.

\section{Availability of data and materials}
The simulation package will be published as open-source software once the paper is accepted.
\\




\section{Funding}
This work is supported by a Career Award at the Scientific Interface from Burroughs Wellcome Fund, the Minnesota Office of Higher Education (award number 257132 / 3000008642), the Research \& Innovation Office at University of Minnesota, and University of Minnesota Graduate School 2024-2025 Doctoral Dissertation Fellowship.\\

\section{Author information}

\noindent \textbf{Authors and Affiliations}\\
Mechanical Engineering, University of Minnesota\\
Daehyun Kim and Jeffrey Tithof\\

\noindent \textbf{Contributions}\\
\noindent Conceptualization: DK, JT; Methodology: DK; Formal analysis and investigation: DK; Writing - original draft preparation: DK; Writing - review and editing: JT; Funding acquisition: DK, JT; Resources: JT; Supervision: JT\\

\noindent \textbf{Corresponding author}\\
\noindent Correspondence to \href{kimx4718@umn.edu}{Daehyun Kim} and \href{tithof\@umn.edu}{Jeffrey Tithof}

\section{Ethics declarations}

\noindent \textbf{Ethical Approval}\\
\noindent Not applicable.\\

\noindent \textbf{Consent for publication}\\
\noindent Not applicable.\\

\noindent \textbf{Competing interests}\\
\noindent The authors declare no competing interests.\\

\end{document}